%% file: main.tex
\documentclass[acmtog,nonacm,screen]{acmart}
\usepackage{multirow}

\usepackage{prettyref}
\usepackage{amsmath} 
\usepackage{tcolorbox} 
\tcbuselibrary{listings, theorems} 

\usepackage[labelfont=bf,format=plain,singlelinecheck=false]{caption}
\usepackage{subcaption}
\usepackage{bm,microtype}
\newcommand{\todo}[1]{\textcolor[RGB]{255, 0, 0}{[\textbf{TODO:} {\em #1}]}}

\definecolor{myblue}{RGB}{31, 119, 180}
\definecolor{myorange}{RGB}{255, 127, 14}
\definecolor{mygreen}{RGB}{44, 160, 44}
\definecolor{myred}{RGB}{214, 39, 40}
\usepackage{soul}
\AtBeginDocument{%
  }

\setcopyright{acmlicensed}

\acmJournal{TOG}

\usepackage{tikz}
\usetikzlibrary{angles, calc}
\tikzset{
  dot/.style={
    circle, fill=black, inner sep=1pt, outer sep=0pt
  },
  dot label/.style={
    circle, inner sep=0pt, outer sep=1pt
  },
  pics/right angle/.append style={
    /tikz/draw, /tikz/angle radius=5pt
  }
}
\usepackage{xcolor}
\usepackage[customcolors]{hf-tikz}

\acmSubmissionID{369}

\usepackage{prettyref}
\newrefformat{fig}{Figure~\ref{#1}}
\newrefformat{par}{Section~\ref{#1}}
\newrefformat{appen}{Appendix~\ref{#1}}
\newrefformat{sec}{Section~\ref{#1}}
\newrefformat{sub}{Section~\ref{#1}}
\newrefformat{table}{Table~\ref{#1}}
\newrefformat{ass}{Assumption~\ref{#1}}
\newrefformat{alg}{Algorithm~\ref{#1}}
\newrefformat{def}{Definition~\ref{#1}}
\newrefformat{thm}{Theorem~\ref{#1}}
\newrefformat{cor}{Corollary~\ref{#1}}
\newrefformat{lem}{Lemma~\ref{#1}}
\newrefformat{step}{Step~\ref{#1}}
\newrefformat{ln}{Line~\ref{#1}}
\newrefformat{rem}{Remark~\ref{#1}}
\newrefformat{eq}{Equation~\ref{#1}}
\newrefformat{pb}{Problem~\ref{#1}}
\newrefformat{it}{Item~\ref{#1}}
\newrefformat{te}{Term~\ref{#1}}
\def\Eqref Eq:#1:{\eqref{eq:#1}}
\newrefformat{Eq}{Equation~\Eqref#1:}

\usepackage{rotating}

\newcommand{\xx}{\mathbf{x}}

\newcommand{\zz}{\mathbf{z}}

\newcommand{\yy}{\mathbf{y}}

\newcommand{\om}{{\mbox{\boldmath{$\omega$}}}}

\newcommand\restr[2]{{\left.\kern-\nulldelimiterspace{}#1\right|_{#2}}}

\newcommand{\ResizedEqNoLabel}[1]{\resizebox{\hsize}{!}{$\begin{aligned}#1\end{aligned}$}}

\usepackage[ruled,linesnumbered]{algorithm2e}
\usepackage{algpseudocode}
\SetAlFnt{\small}
\SetAlCapFnt{\small}
\SetAlCapNameFnt{\small}
\SetAlCapHSkip{0pt}

\usepackage{svg}
\usepackage{xcolor}
\usepackage{wrapfig}
\usepackage[export]{adjustbox}
\usepackage{colortbl}

\begin{document}

\title{Real-time Neural Six-way Lightmaps}

\author{Wei Li}
\authornote{Both authors contributed equally to this research.}
\orcid{0000-0001-8754-6679}
\email{1104720604wei@gmail.com}
\affiliation{%
  \institution{Shanghai Jiao Tong University}
    \city{Shanghai}
  \country{China}
}

\author{Hanxiao Sun}
\authornotemark[1]
\orcid{0009-0009-8887-2148}
\email{hx.sun@mail.nankai.edu.cn}
\affiliation{%
  \institution{LIGHTSPEED}
    \city{Shenzhen}
  \country{China}
}

\author{Tao Huang}
\orcid{0009-0002-3458-0851}
\email{cstgcaiji@gmail.com}
\affiliation{%
  \institution{LIGHTSPEED}
    \city{Shenzhen}
  \state{GuangDong}
  \country{China}
}

\author{Haoxiang Wang}
\orcid{0000-0001-7255-4641}
\email{whx22@mails.tsinghua.edu.cn}
\affiliation{%
  \institution{Tsinghua University}
    \city{Beijing}
  \country{China}
}

\author{Tongtong Wang}
\orcid{0009-0005-6585-3009}
\email{wangtong923@gmail.com}
\affiliation{%
  \institution{LIGHTSPEED}
    \city{Sydney}
  \country{Australia}
}

\author{Zherong Pan}
\orcid{0000-0001-9348-526X}
\email{zherong.pan.usa@gmail.edu}
\affiliation{%
  \institution{LIGHTSPEED}
    \city{Seattle}
  \state{WA}
  \country{USA}
}

\author{Kui Wu}
\orcid{0000-0003-3326-7943}
\email{walker.kui.wu@gmail.com}
\affiliation{%
  \institution{LIGHTSPEED}
    \city{Los Angeles}
  \state{CA}
  \country{USA}
}

\begin{abstract}
Participating media are a pervasive and intriguing visual effect in virtual environments. Unfortunately, rendering such phenomena in real-time is notoriously difficult due to the computational expense of estimating the volume rendering equation. While the six-way lightmaps technique has been widely used in video games to render smoke with a camera-oriented billboard and approximate lighting effects using six precomputed lightmaps, achieving a balance between realism and efficiency, it is limited to pre-simulated animation sequences and is ignorant of camera movement. In this work, we propose a neural six-way lightmaps method to strike a long-sought balance between dynamics and visual realism. Our approach first generates a guiding map from the camera view using ray marching with a large sampling distance to approximate smoke scattering and silhouette. Then, given a guiding map, we train a neural network to predict the corresponding six-way lightmaps. The resulting lightmaps can be seamlessly used in existing game engine pipelines. This approach supports visually appealing rendering effects while enabling real-time user interactivity, including smoke-obstacle interaction, camera movement, and light change. By conducting a series of comprehensive benchmarks, we demonstrate that our method is well-suited for real-time applications, such as games and VR/AR.
\end{abstract}


%
%


\begin{teaserfigure}
\newcommand{\figcap}[1]{\begin{minipage}{0.249\linewidth}\centering#1\end{minipage}}
\rotatebox{90}{\hspace*{.07\linewidth}Frame 0 }\hfill%
\includegraphics[trim = 0 550 0 600, clip, width=0.98\textwidth]{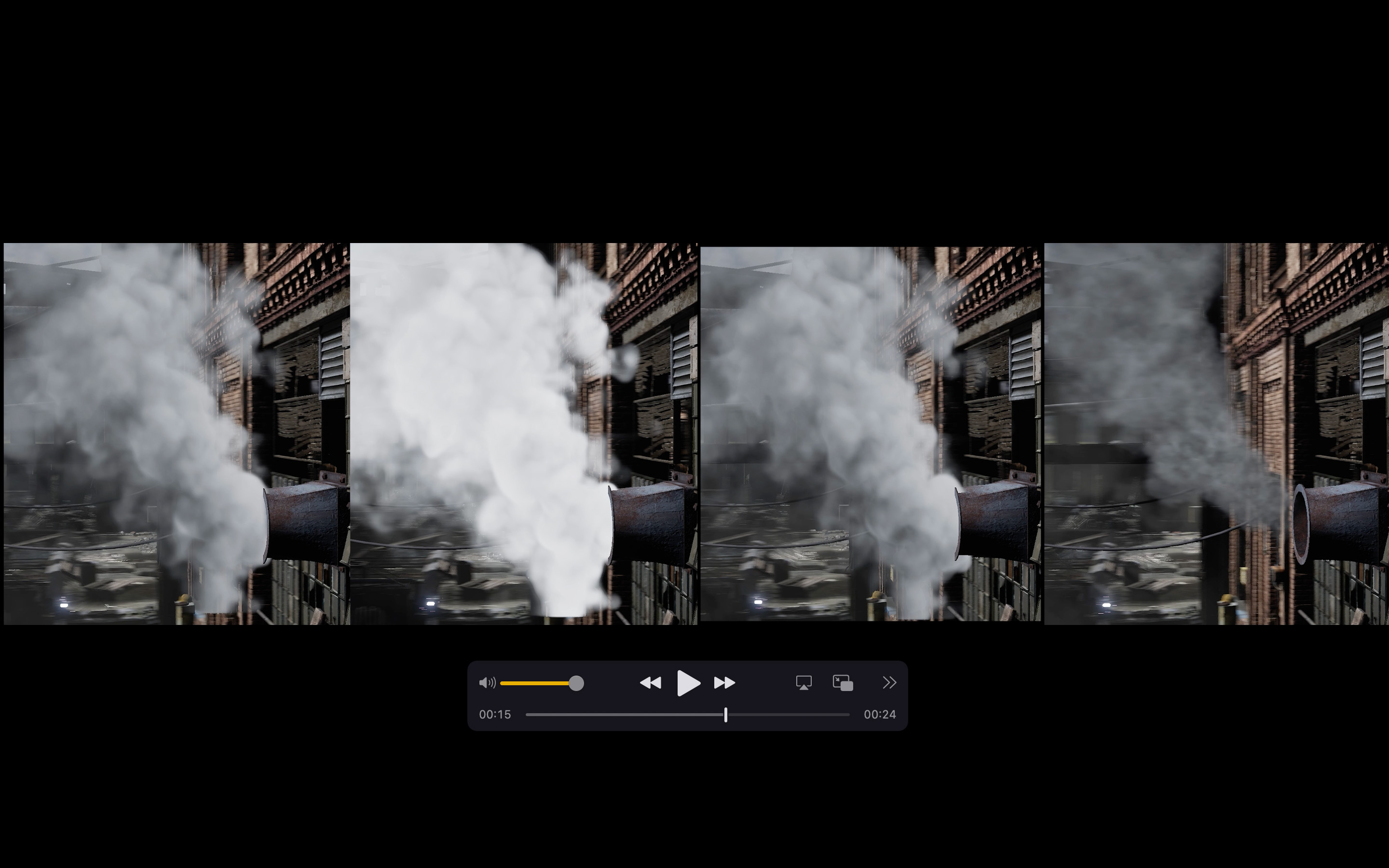} \\
\rotatebox{90}{\hspace*{.07\linewidth}Frame 200}\hfill%
\includegraphics[trim = 0 550 0 600, clip, width=0.98\textwidth]{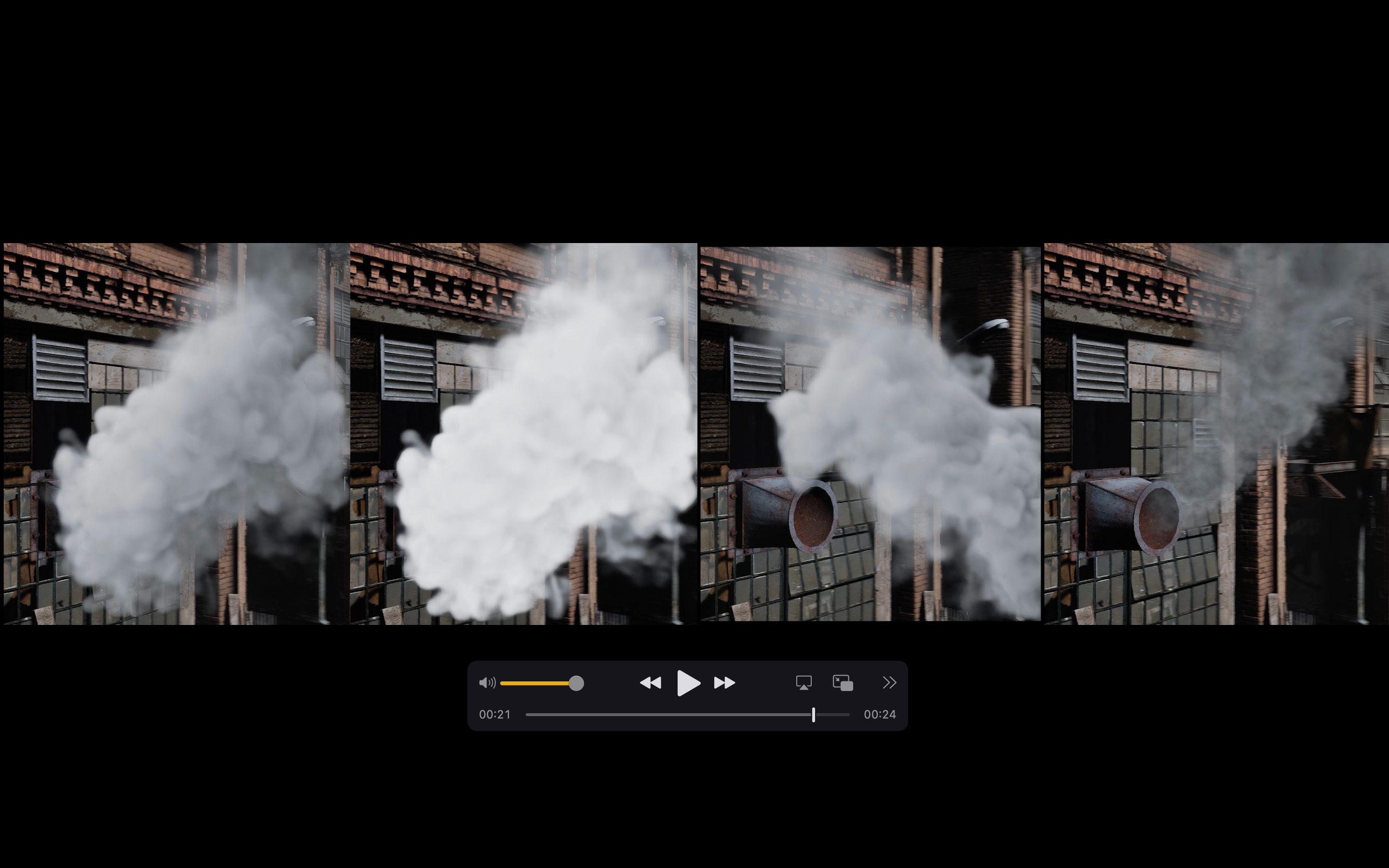}
\figcap{Ours}\hfill
\figcap{W/o lightmaps}\hfill
\figcap{Traditional six-way lightmaps}\hfill
\figcap{Niagara system in Unreal Engine} \vspace{-1.em}\\
\caption
{Two frames of a chimney smoke example rendered with rotating camera direction in Unreal Engine (UE)~\cite{unrealengine}. Our neural lightmaps support dynamic lighting, camera direction, and realistic multiple-scattering effects, whereas traditional six-way lightmaps~\cite{Muller2023sixway} with asymmetrically shaped smoke only work for one camera direction, leading to severe offset artifacts and a lack of interaction. Compared with UE built-in Niagara system using sprite-based particle effects, our method produces substantially more realistic smoke appearance.} 
\label{fig:teaser}
\Description{}
\end{teaserfigure}


\maketitle
\input{intro.tex}
\input{related.tex}
\input{background.tex}
\input{method.tex}
\input{result.tex}
\input{conclusion.tex}

\bibliographystyle{ACM-Reference-Format}
\bibliography{bibliography}

\clearpage
\newpage

\begin{figure}[ht!]
\centering
\newcommand{\figcap}[1]{\begin{minipage}{0.249\linewidth}\centering#1\end{minipage}}
\figcap{Reference}\hfill
\figcap{Ours}\hfill
\figcap{ReSTIR$^*$}\hfill
\figcap{MRPNN} \vspace{-1em}\\
\includegraphics[trim = 0 30 0 80, clip, width=0.249\linewidth]{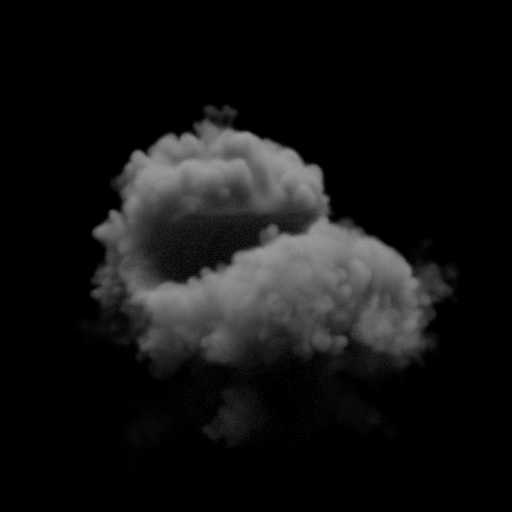}\hfill
\includegraphics[trim = 0 30 0 80, clip, width=0.249\linewidth]{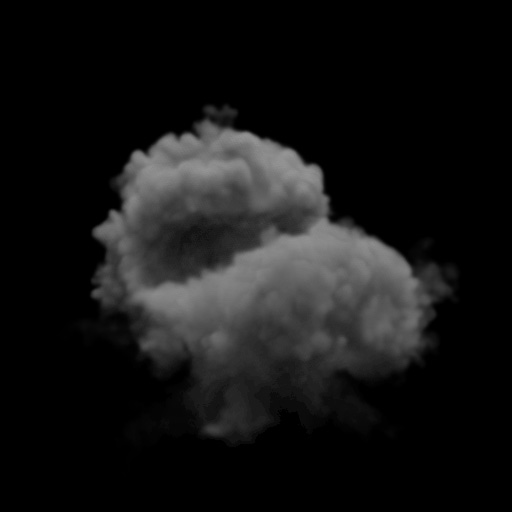}\hfill
\includegraphics[trim = 0 30 0 80, clip, width=0.249\linewidth]{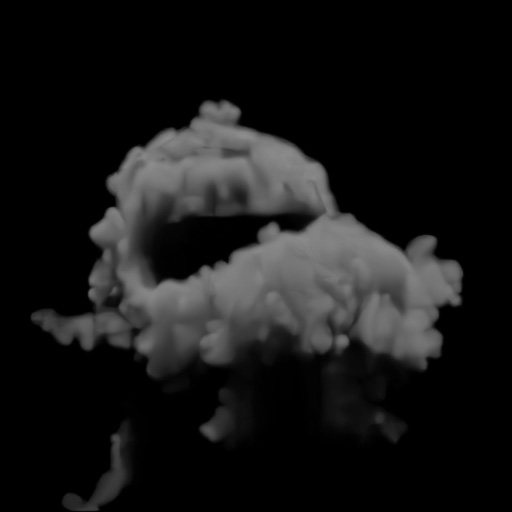}\hfill
\includegraphics[trim = 0 30 0 80, clip, width=0.249\linewidth]{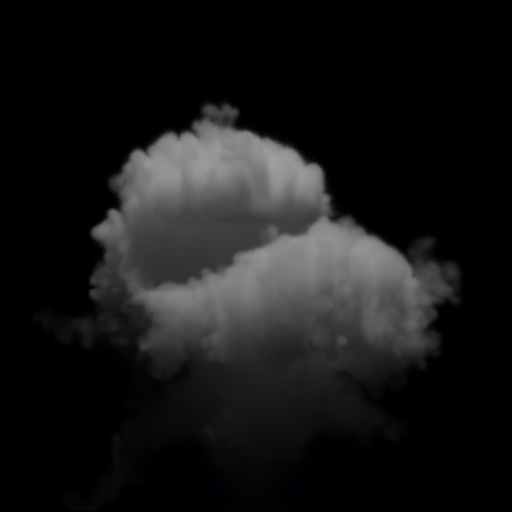}\\
\figcap{PSNR $\uparrow$ / MSE $\downarrow$}\hfill
\figcap{\textbf{35.89}/\textbf{0.00025}}\hfill
\figcap{31.55/0.00072}\hfill
\figcap{35.23/0.00030} 
\caption{\label{fig:density100} {Comparison on a denser smoke field shows that our method continues to outperform prior techniques, maintaining higher visual fidelity even under significantly increased density.}}
\Description{}
\end{figure} 

\begin{figure}[ht!]
\centering
\newcommand{\figcap}[1]{\begin{minipage}{0.249\linewidth}\centering#1\end{minipage}}
\figcap{Reference}\hfill
\figcap{Front}\hfill
\figcap{Front+LR}\hfill
\figcap{Front+TB} \vspace{-1em}\\
\includegraphics[trim = 0 30 0 80, clip, width=0.249\linewidth]{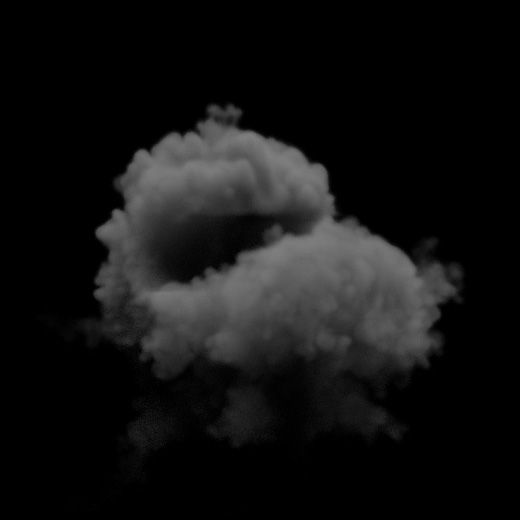}\hfill
\includegraphics[trim = 0 20 0 60, clip, width=0.249\linewidth]{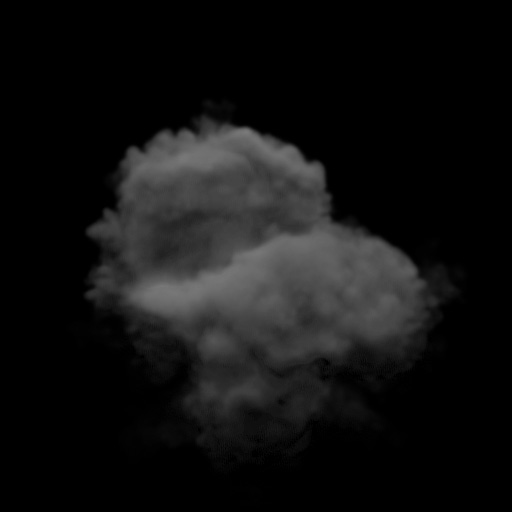}\hfill
\includegraphics[trim = 0 20 0 60, clip, width=0.249\linewidth]{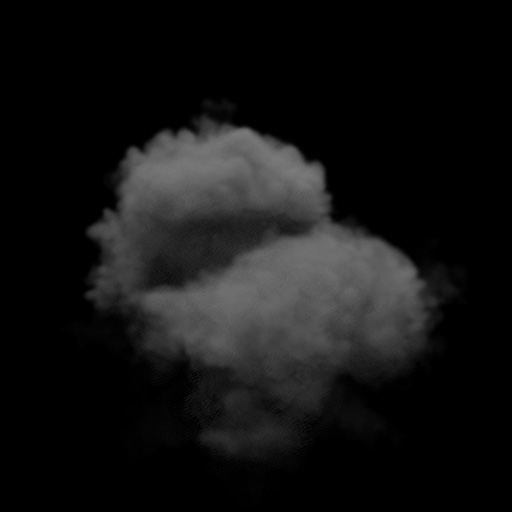}\hfill
\includegraphics[trim = 0 30 0 80, clip, width=0.249\linewidth]{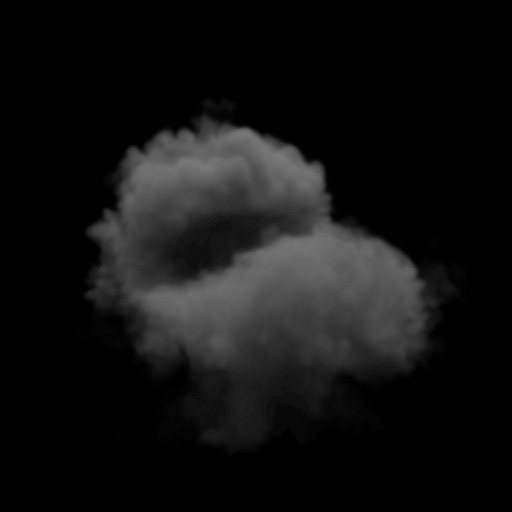}\\
\figcap{\small PSNR$\uparrow$ }\hfill
\figcap{\small 32.61/37.83/28.96}\hfill
\figcap{\small 34.99/39.05/33.08}\hfill
\figcap{\small \textbf{40.85/48.78/37.93}}\\
\figcap{\small MSE $\downarrow$ }\hfill
\figcap{\scriptsize 0.00065/0.00127/0.00017}\hfill
\figcap{\scriptsize 0.00033/0.00049/0.00012}\hfill
\figcap{\scriptsize \textbf{0.00009/0.00016/0.00001}}
\caption{\label{fig:directionstudy} {Comparison on different illumination configurations for guiding map generation with Avg./max/min PSNR and MSE. LR and TB denote left + right and top + bottom, respectively.}}
\Description{}
\end{figure}

\begin{figure} [ht!]
\centering
\newcommand{\figcap}[1]{\begin{minipage}{0.245\linewidth}\centering#1\end{minipage}}
\includegraphics[trim = 620 0 620 0, clip, width=0.249\linewidth]{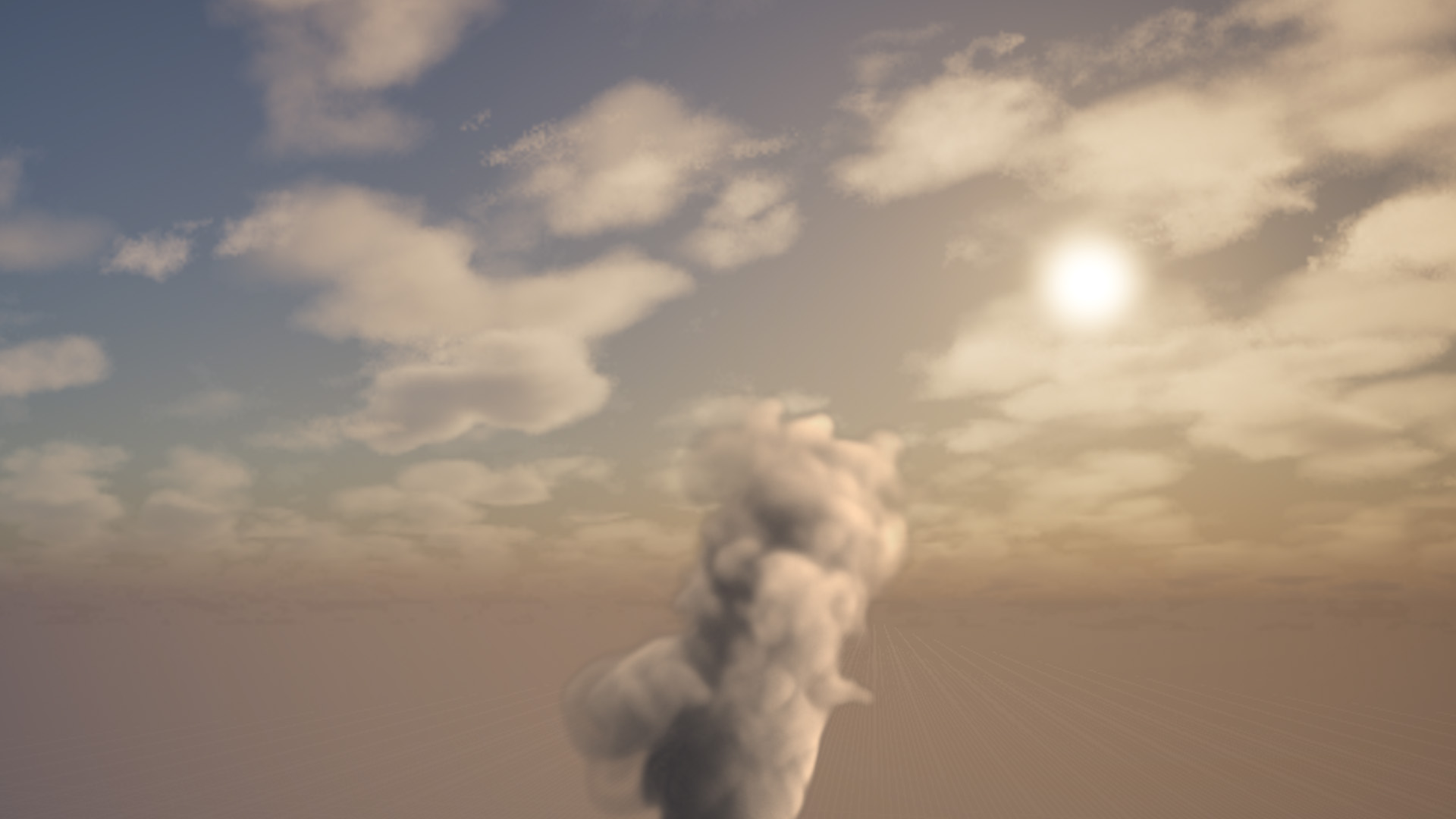}\hfill
\includegraphics[trim = 620 0 620 0, clip, width=0.249\linewidth]{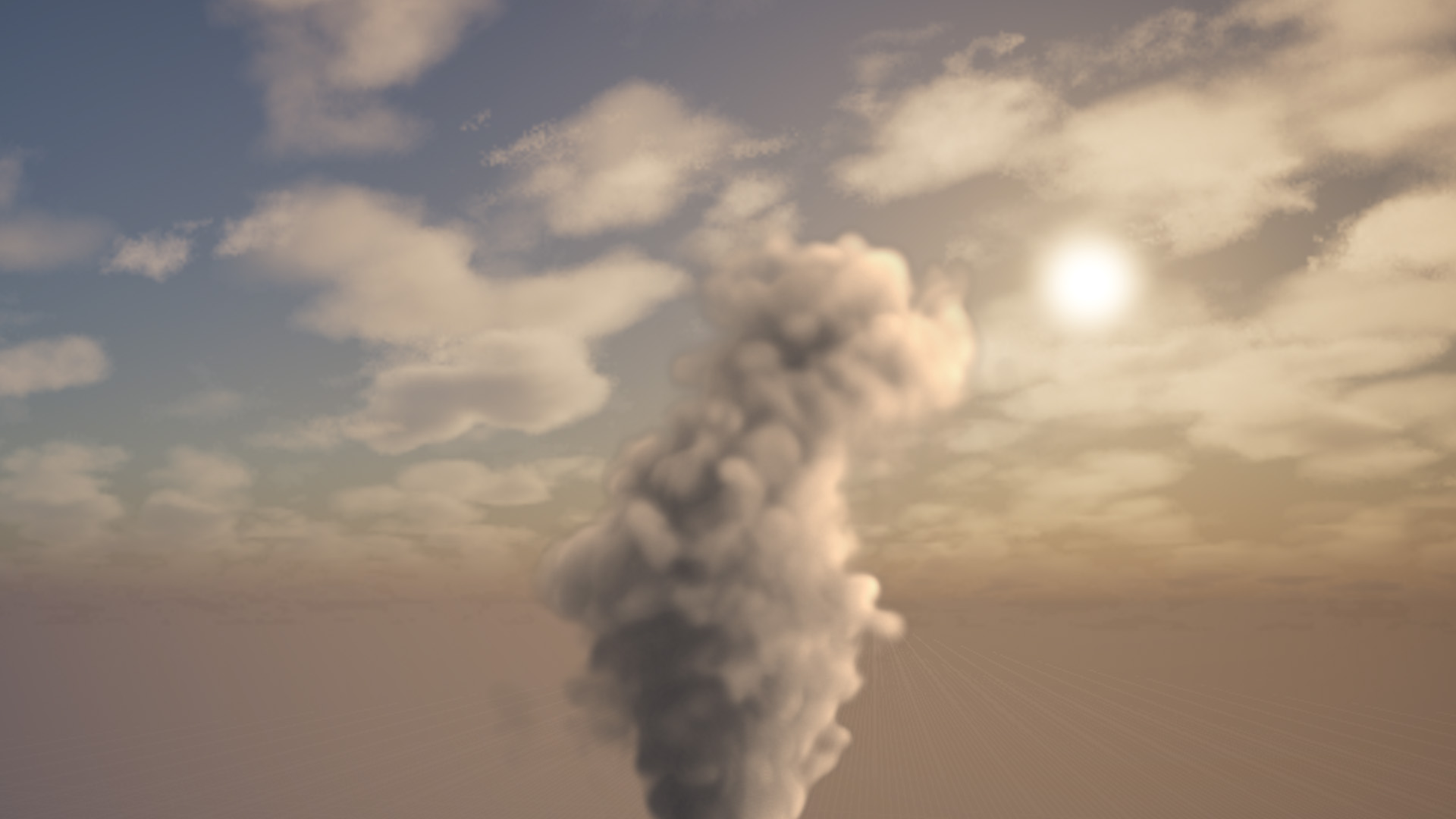}\hfill
\includegraphics[trim = 620 0 620 0, clip, width=0.249\linewidth]{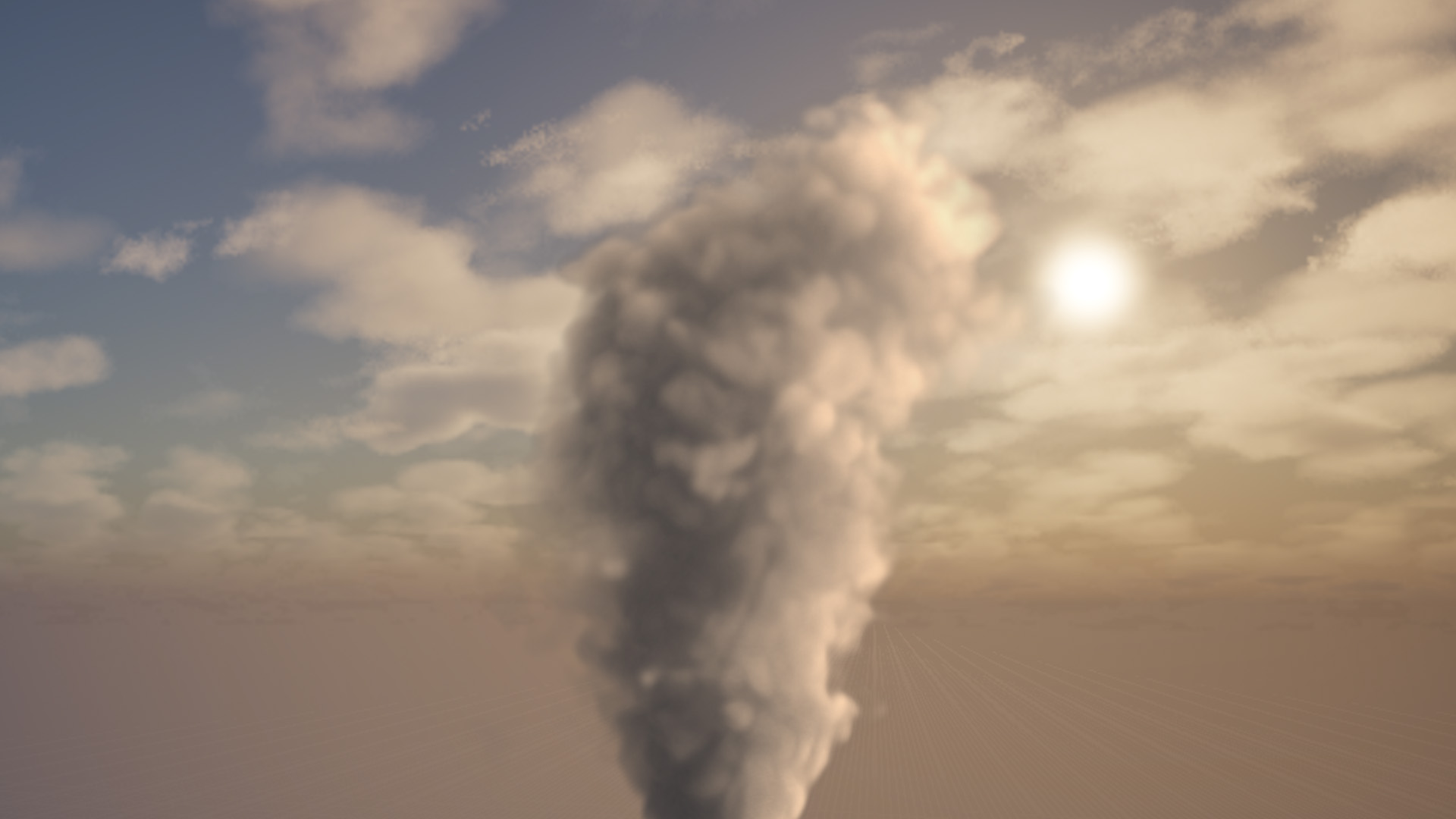}\hfill
\includegraphics[trim = 620 0 620 0, clip, width=0.249\linewidth]{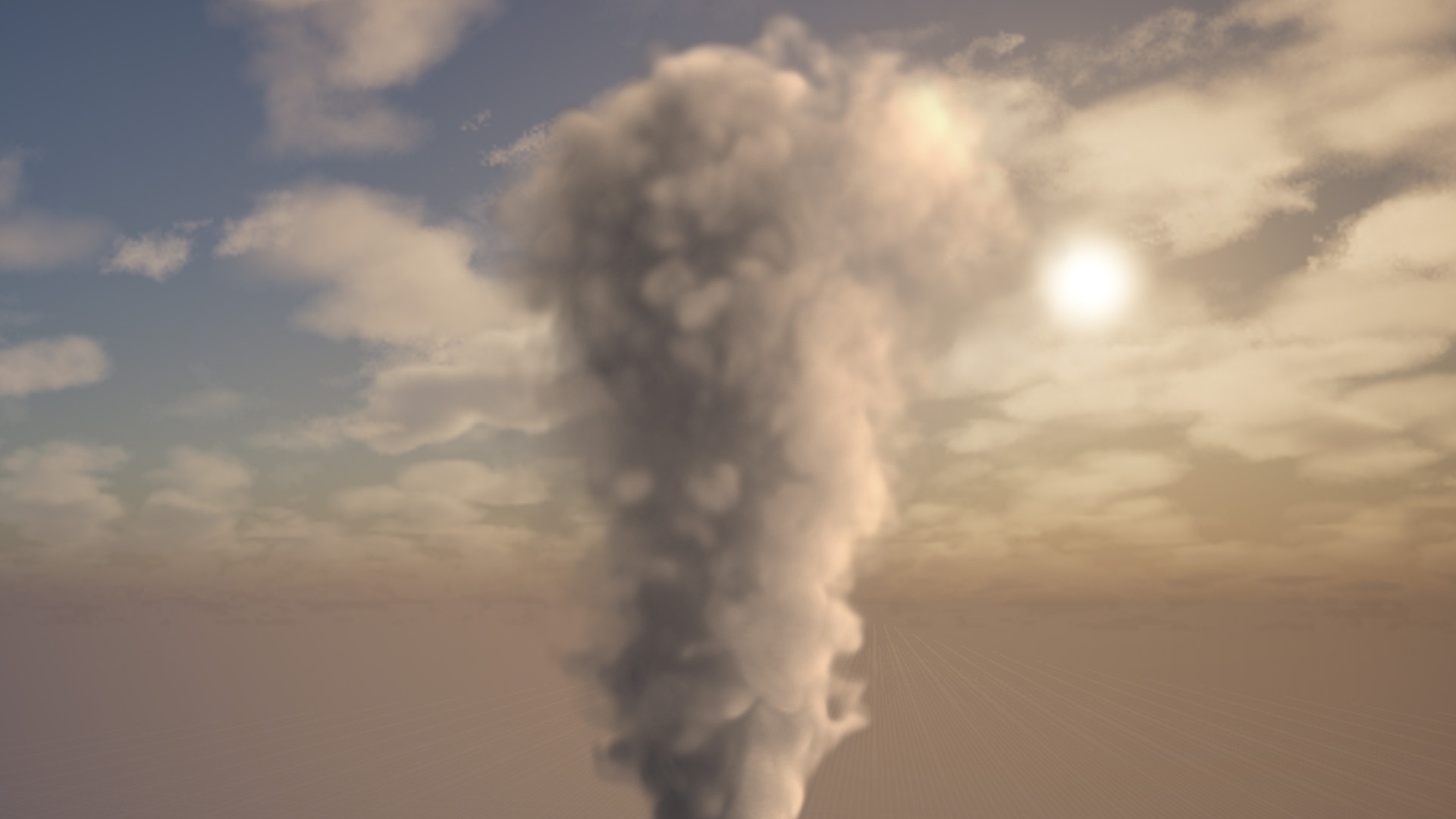}
\caption{\label{fig:jetflow} {A rotating jet flow, where our method can illuminate smoke wit details under the shadow.}}
\Description{}
\end{figure} 

\begin{figure} [ht!]
\centering
\includegraphics[trim = 0 70 0 70, clip, width=0.499\linewidth]{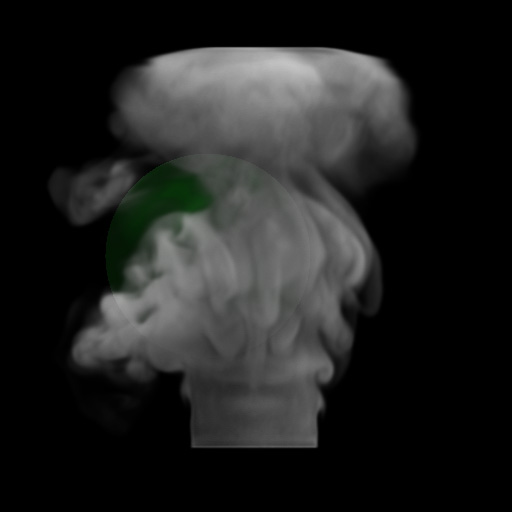}\hfill
\includegraphics[trim = 0 70 0 70, clip, width=0.499\linewidth]{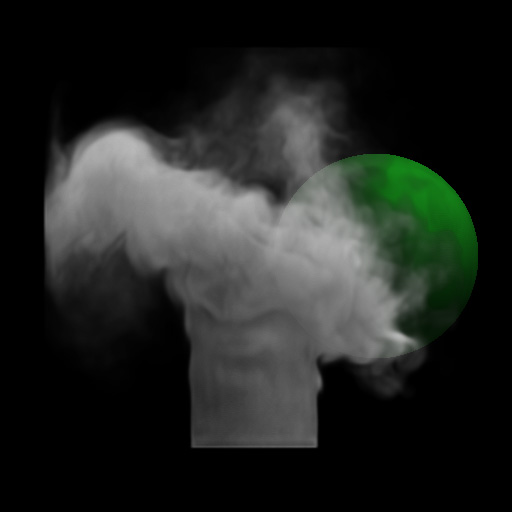}
\caption{\label{fig:sphere} {Jet flow over a moving sphere, demonstrating the rich smoke details illuminated by our method.}}
\Description{}
\end{figure} 

\begin{figure} [hb!]
\centering
\includegraphics[trim = 650 40 800 300, clip, width=0.249\linewidth]{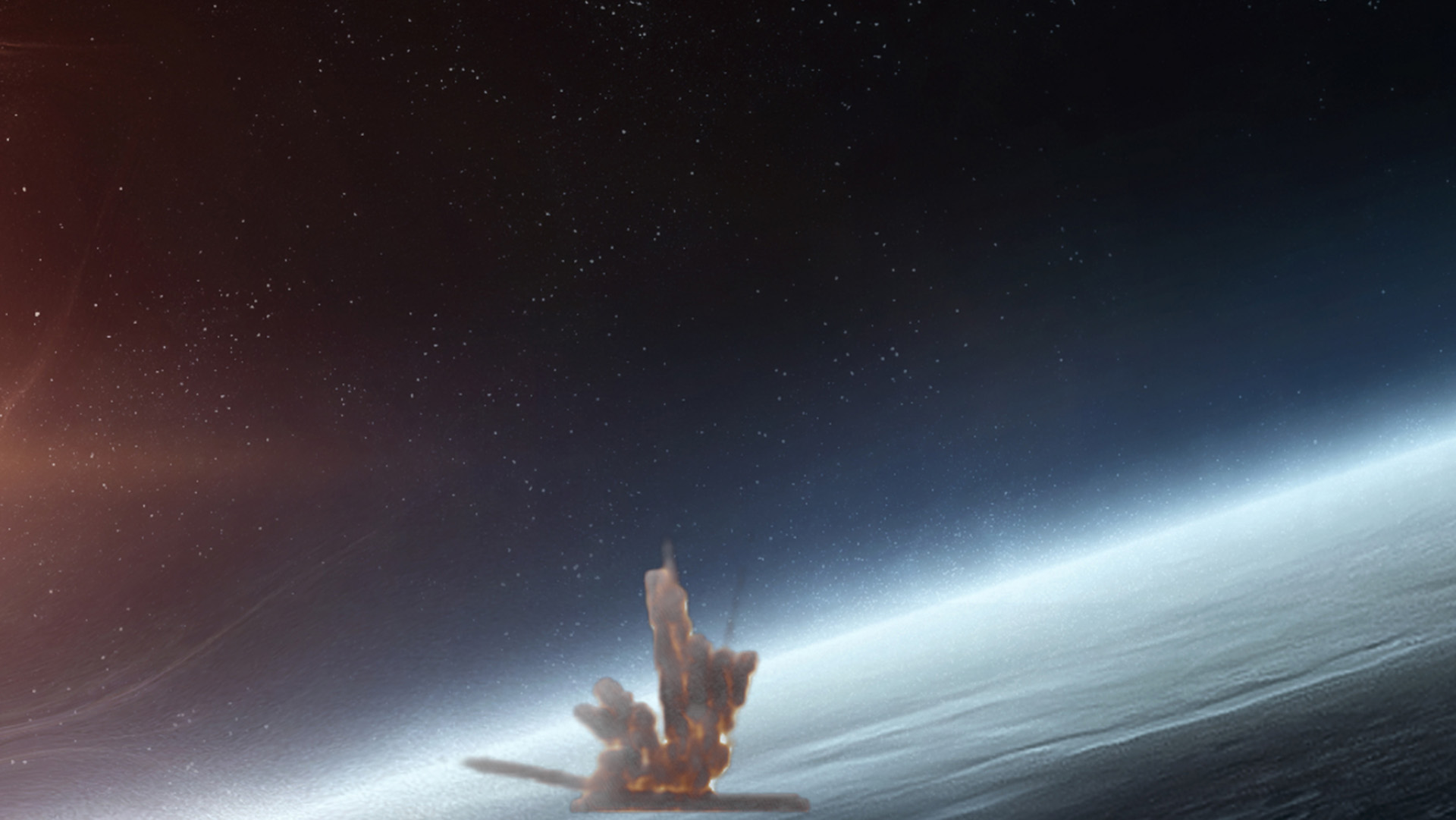}\hfill
\includegraphics[trim = 650 40 800 300, clip, width=0.249\linewidth]{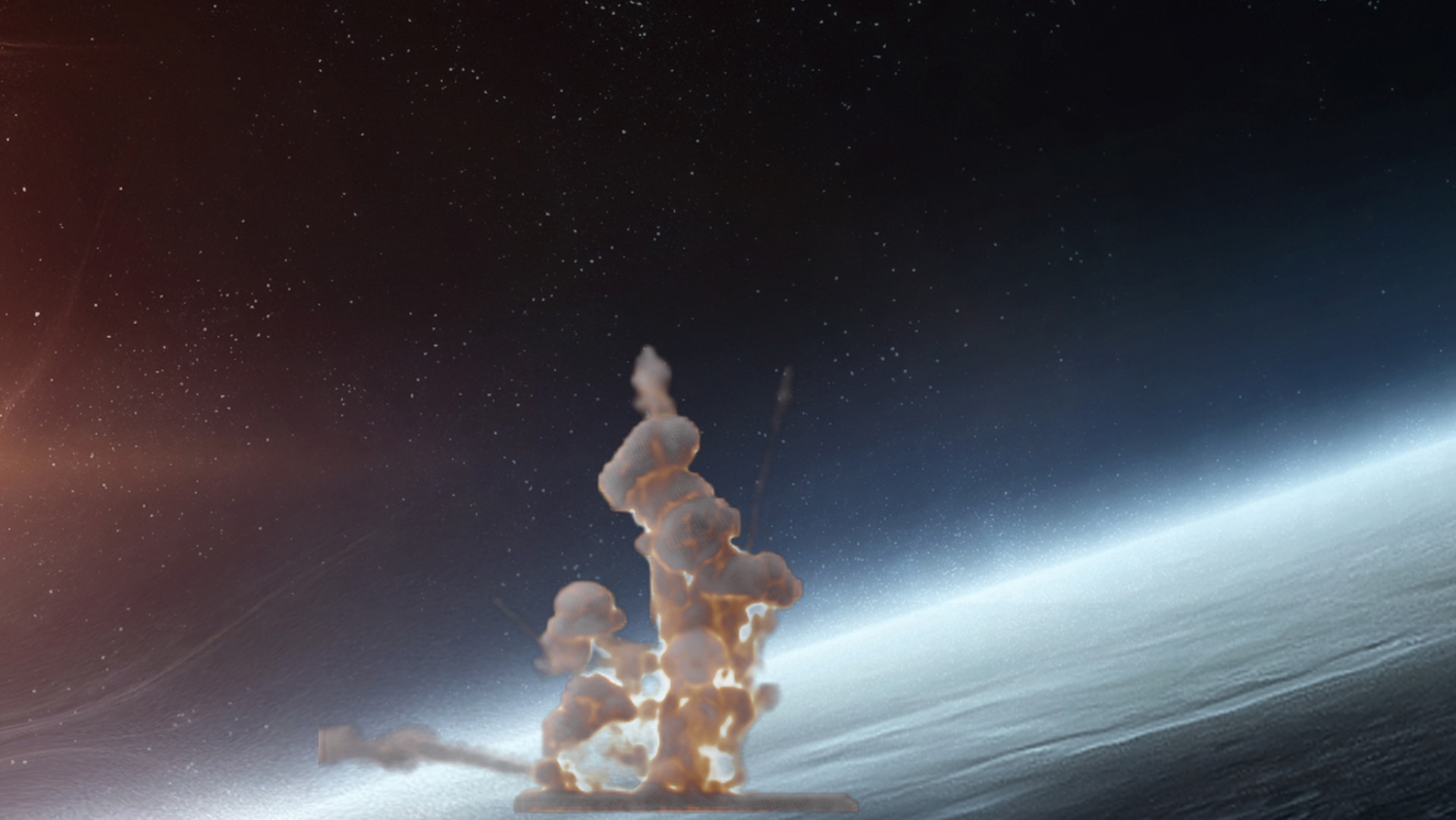}\hfill
\includegraphics[trim = 650 40 800 300, clip, width=0.249\linewidth]{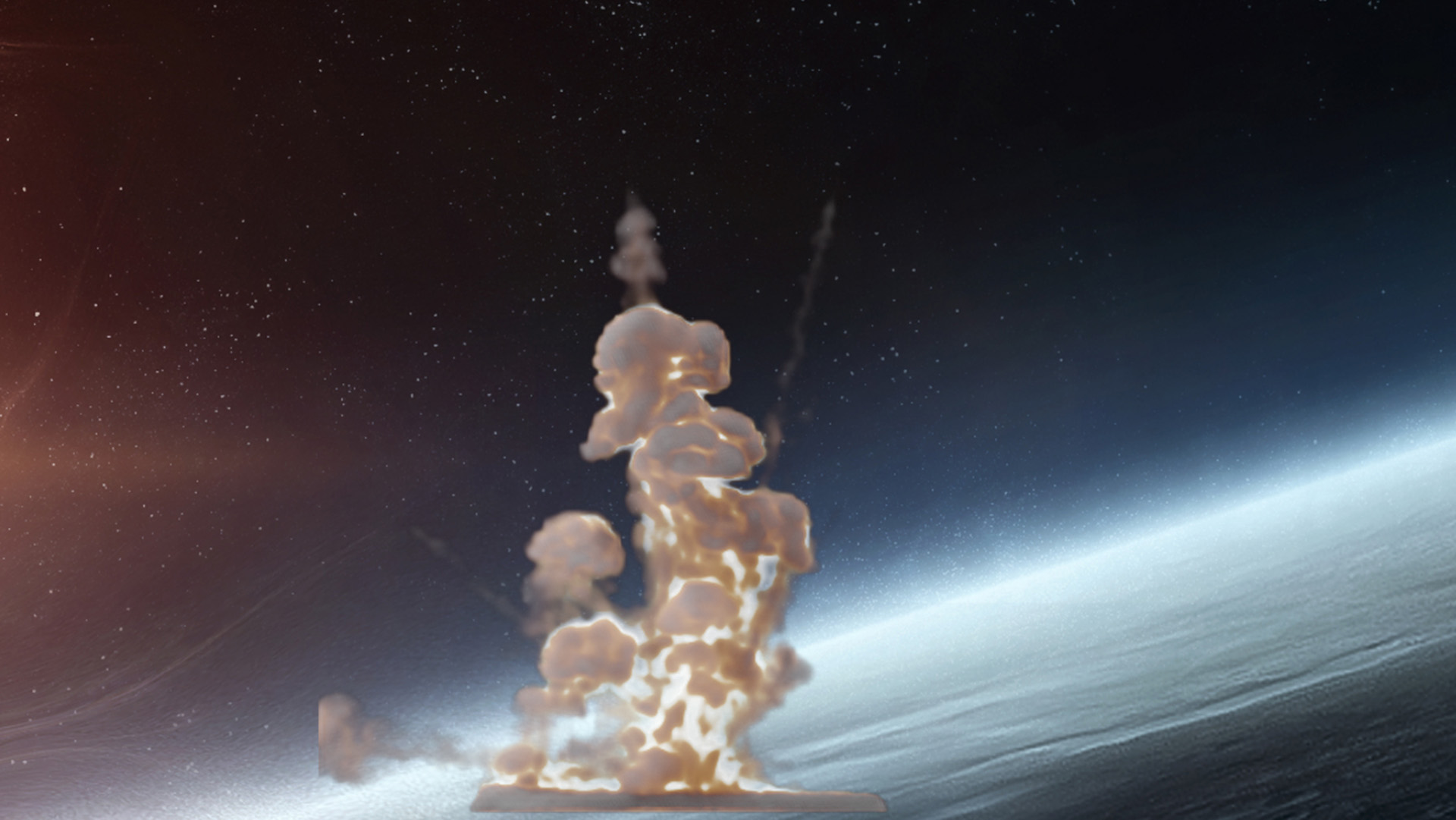}\hfill
\includegraphics[trim = 650 40 800 300, clip, width=0.249\linewidth]{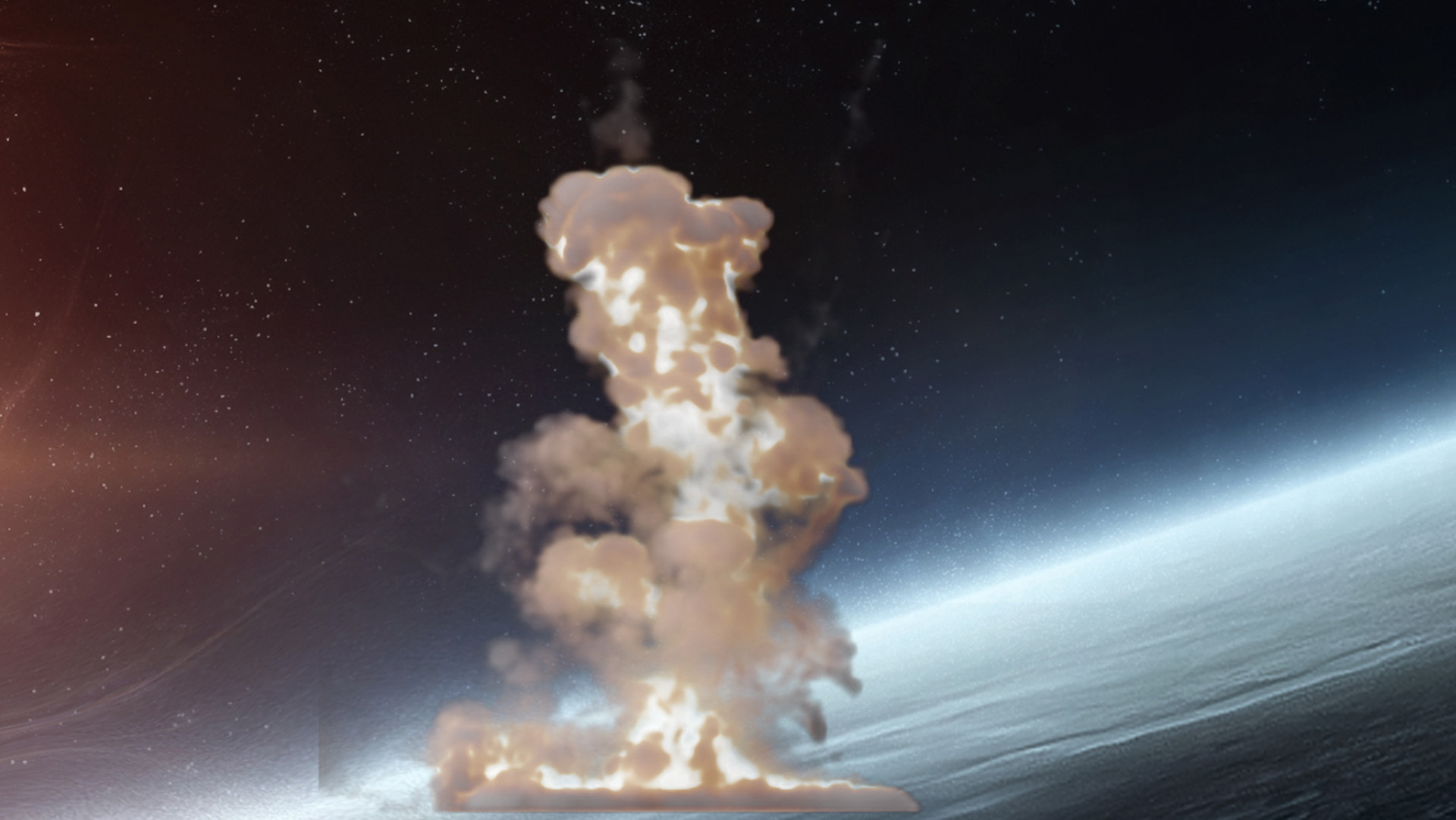}\hfill
\caption{\label{fig:explosion} {Explosions from the third-party simulation result, and the emissive term is inferred from our method to illuminate the high-temperature area. }}
\Description{}
\end{figure} 

\begin{figure} [hb!]
\centering
\includegraphics[trim = 0 0 0 150, clip,width=0.499\columnwidth]{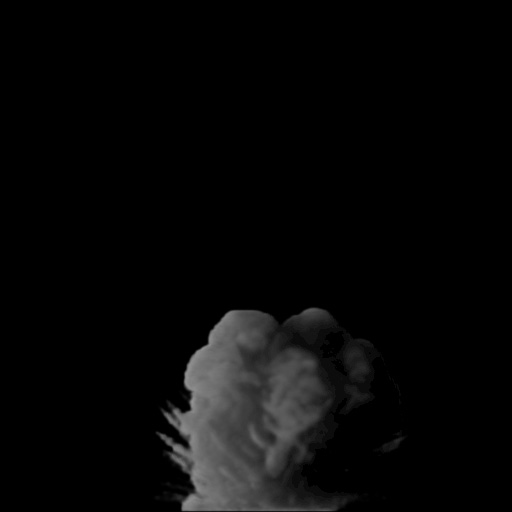}\hfill
\includegraphics[trim = 0 0 0 150, clip,width=0.499\columnwidth]{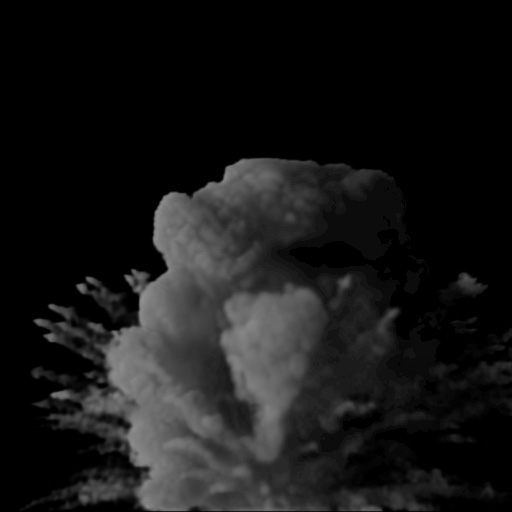}
\caption{\label{fig:third} Our method can illuminate pre-simulated explosion data from the third-party website directly.}
\Description{}
\end{figure} 

\begin{figure} [hb!]
\centering
\includegraphics[trim = 600 200 600 400, clip, width=0.495\linewidth]{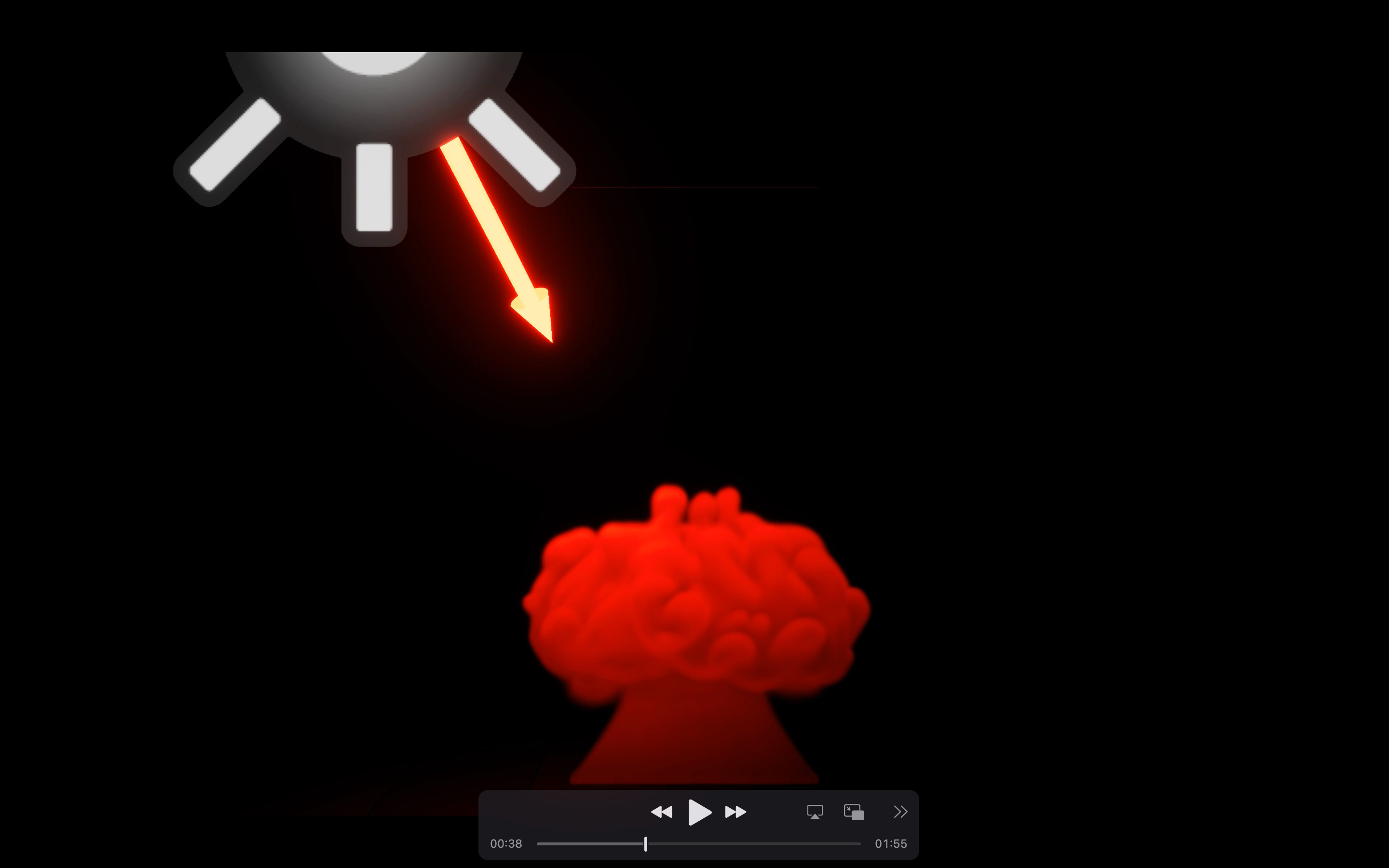} \hfill
\includegraphics[trim = 600 200 600 400, clip, width=0.495\linewidth]{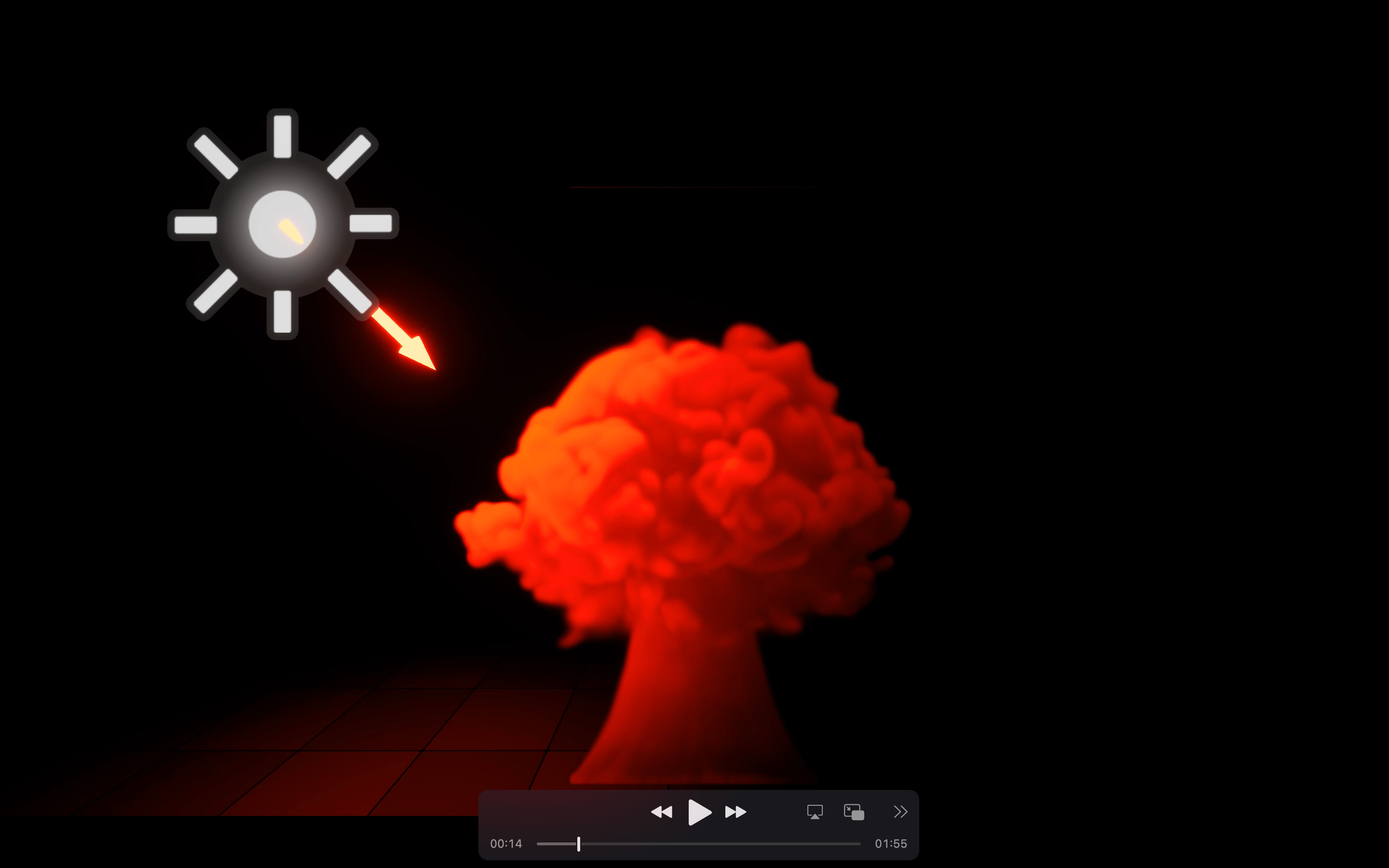} \\ 
\includegraphics[trim = 600 200 600 400, clip, width=0.495\linewidth]{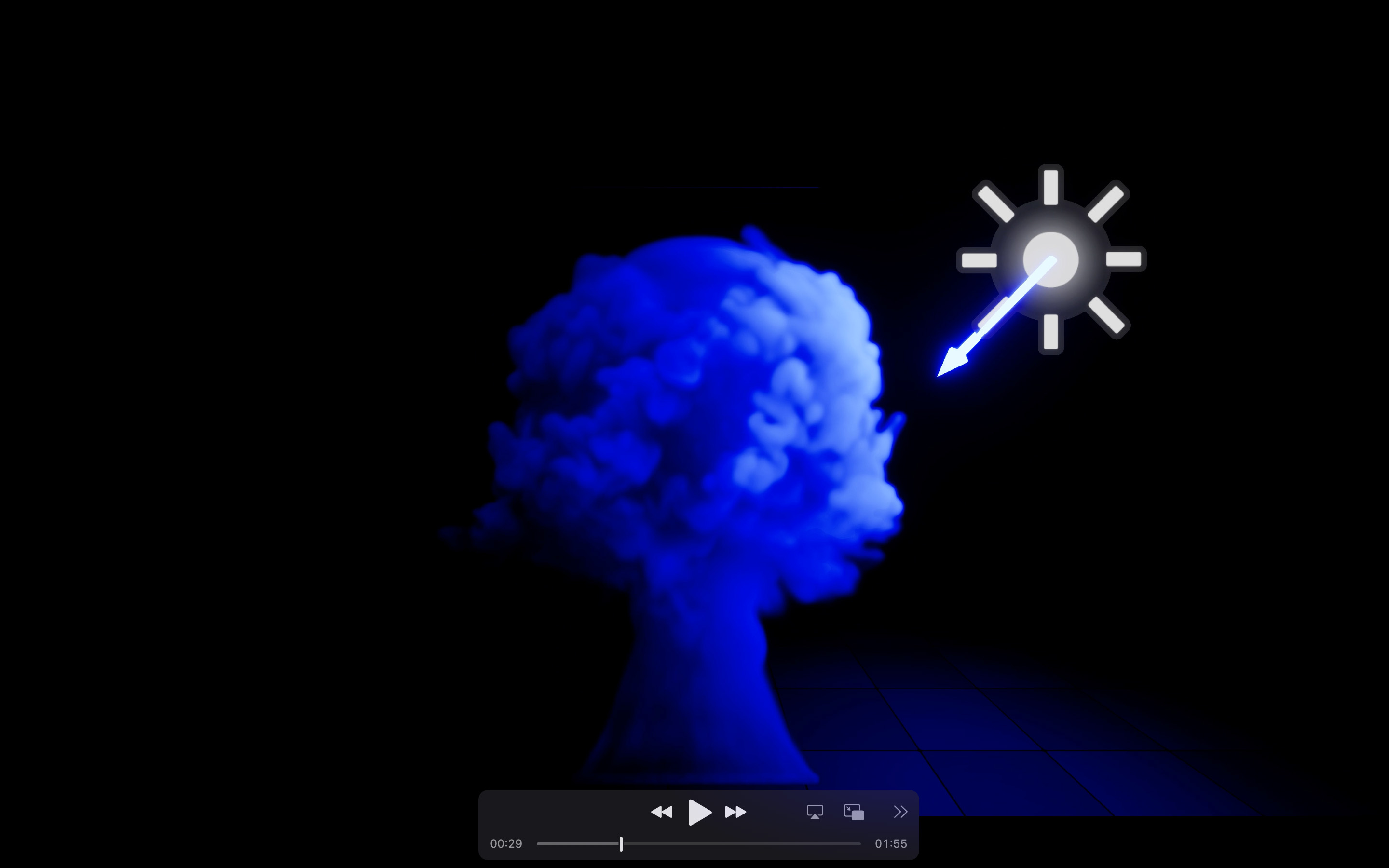} \hfill
\includegraphics[trim = 600 200 600 400, clip, width=0.495\linewidth]{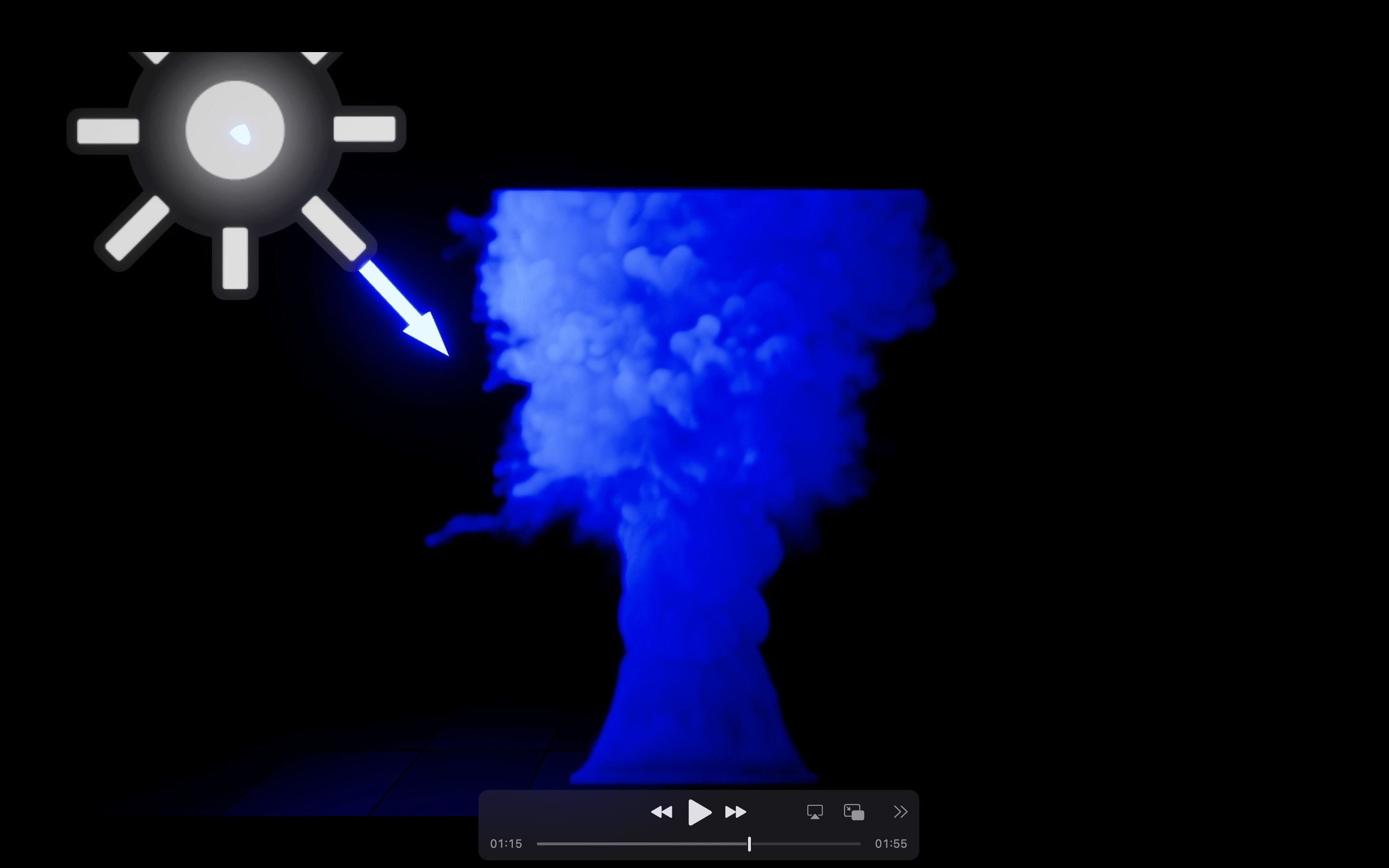} \\ 
\caption{\label{fig:interactive} Our method supports dynamically adjusting light color, position, and orientation.}
\Description{}
\end{figure} 

\begin{figure}[hb!]
\centering
\newcommand{\figcap}[1]{\begin{minipage}{0.249\linewidth}\centering#1\end{minipage}}
\figcap{Reference}\hfill
\figcap{Ours}\hfill
\figcap{ReSTIR$^*$}\hfill
\figcap{MRPNN} \vspace{-1em}\\
\includegraphics[trim = 0 0 0 0, clip, width=0.249\linewidth]{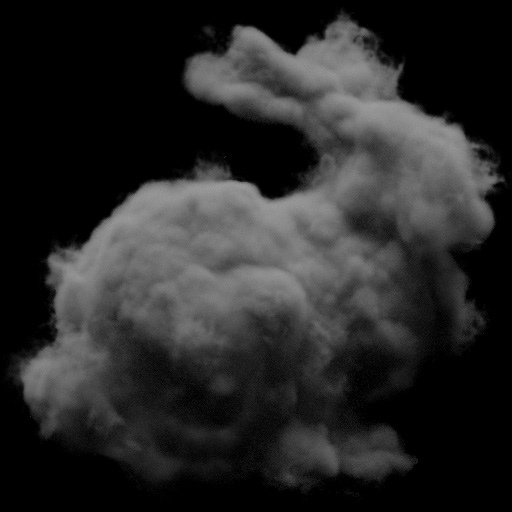}\hfill
\includegraphics[trim = 0 0 0 0, clip, width=0.249\linewidth]{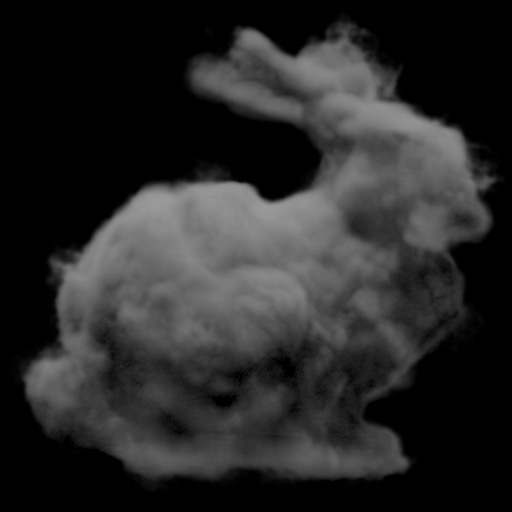}\hfill
\includegraphics[trim = 0 0 0 0, clip, width=0.249\linewidth]{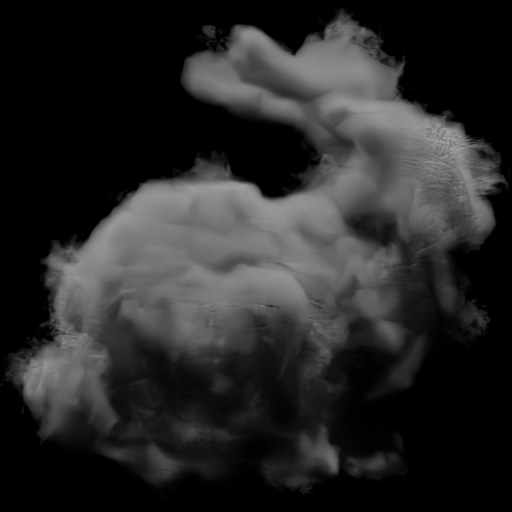}\hfill
\includegraphics[trim = 0 0 0 0, clip, width=0.249\linewidth]{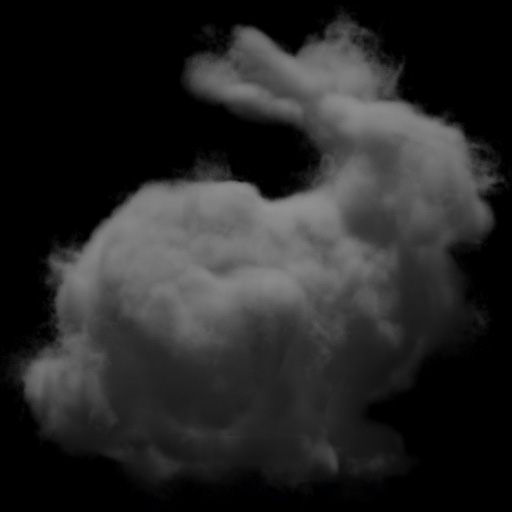}\\
\figcap{PSNR $\uparrow$ / MSE $\downarrow$}\hfill
\figcap{33.44/0.00045}\hfill
\figcap{\textbf{37.88}/\textbf{0.00016}}\hfill
\figcap{32.86/0.00051} 
\caption{\label{fig:vdb_bunny} Rendering a volumetric bunny with ours, ReSTIR~\cite{Lin2021}, and MRPNN~\cite{Hu2023}. While our result is closer to the reference, it still misses the strong shadowing underneath. }
\Description{}
\end{figure}

\begin{figure*}[ht!]
\centering
\newcommand{\figcap}[1]{\begin{minipage}{0.199\linewidth}\centering#1\end{minipage}}
\figcap{Reference}\hfill
\figcap{Ours}\hfill
\figcap{ReSTIR}\hfill
\figcap{ReSTIR$^*$}\hfill
\figcap{MRPNN} \vspace{0.25em}\\
\includegraphics[trim = 0 40 0 80, clip, width=0.199\linewidth]{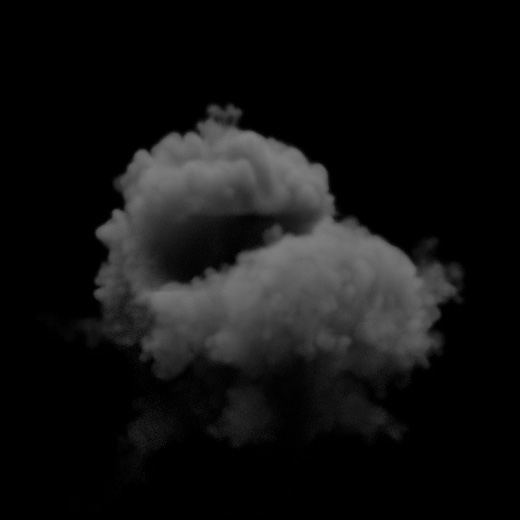}\hfill
\includegraphics[trim = 0 40 0 80, clip, width=0.199\linewidth]{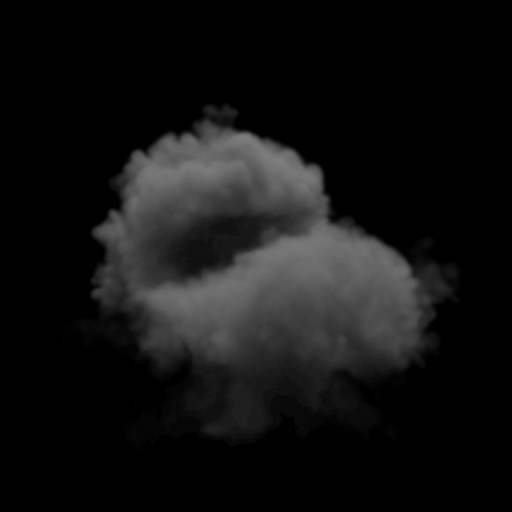}\hfill
\includegraphics[trim = 0 40 0 80, clip, width=0.199\linewidth]{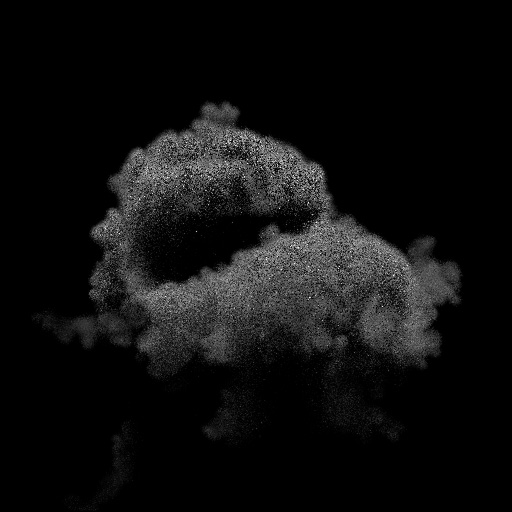}\hfill
\includegraphics[trim = 0 40 0 80, clip, width=0.199\linewidth]{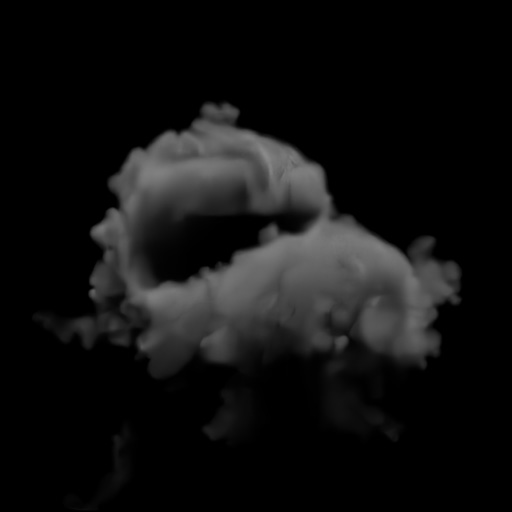}\hfill
\includegraphics[trim = 0 40 0 80, clip, width=0.199\linewidth]{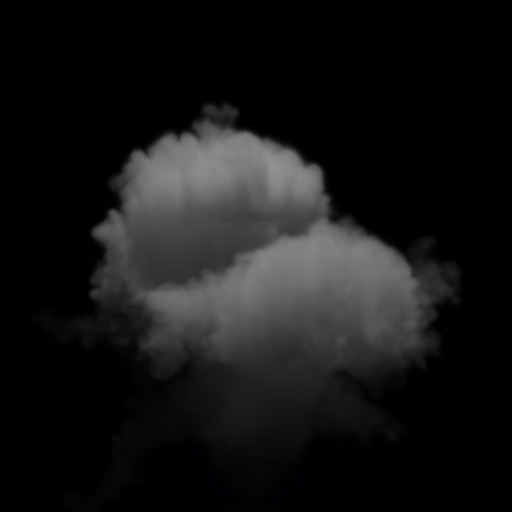}\\
\figcap{PSNR $\uparrow$ / MSE $\downarrow$, Time $\downarrow$}\hfill
\figcap{\textbf{40.71}/\textbf{0.00008}, \textbf{3.9} ms}\hfill
\figcap{26.26/0.00109, 10.4 ms}\hfill
\figcap{34.84/0.00236, 12.6 ms}\hfill
\figcap{36.13/0.00032, 4.0 + 180 ms} \vspace{0.25em} \\
\includegraphics[trim = 0 40 0 0, clip, width=0.199\linewidth]{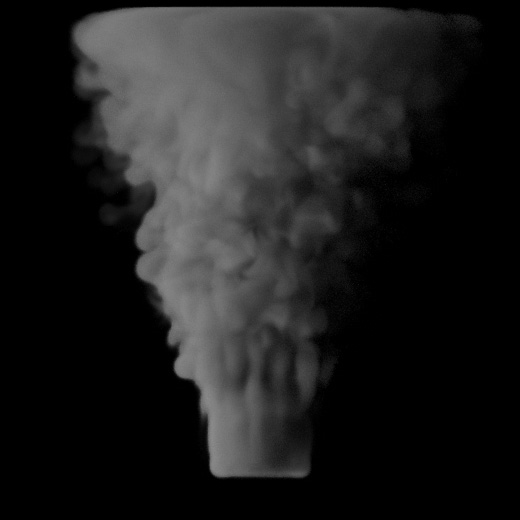}\hfill
\includegraphics[trim = 0 40 0 0, clip, width=0.199\linewidth]{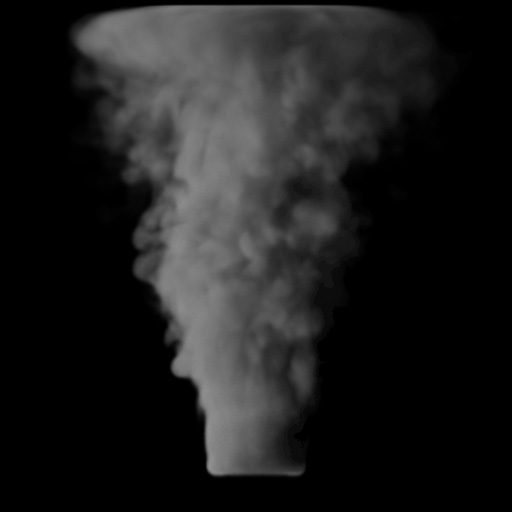}\hfill
\includegraphics[trim = 0 40 0 0, clip, width=0.199\linewidth]{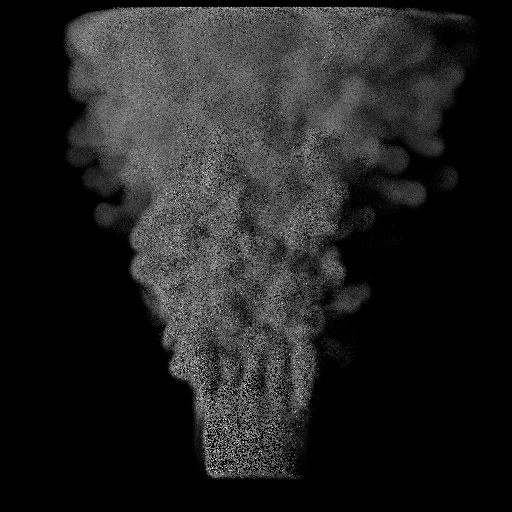}\hfill
\includegraphics[trim = 0 40 0 0, clip, width=0.199\linewidth]{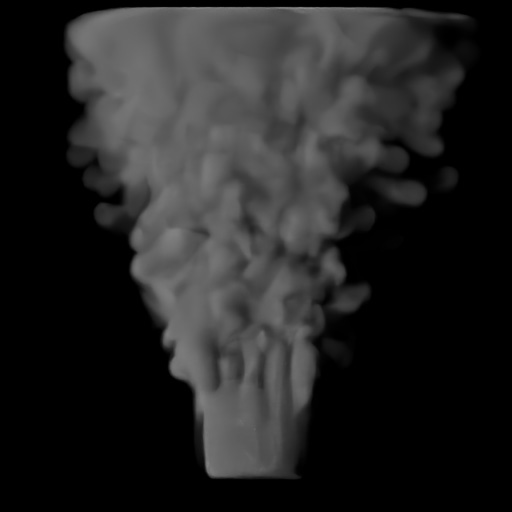}\hfill
\includegraphics[trim = 0 40 0 0, clip, width=0.199\linewidth]{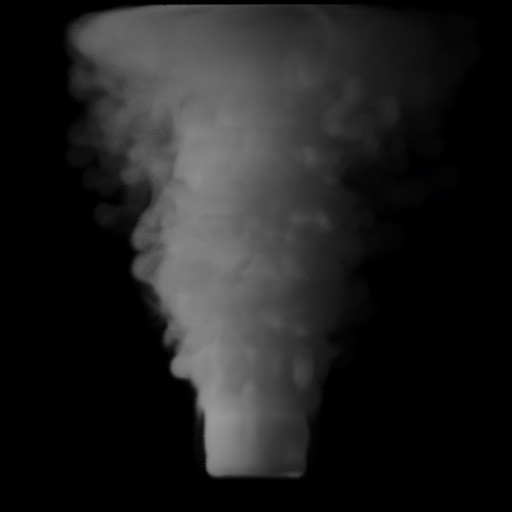}\\
\figcap{PSNR $\uparrow$ / MSE $\downarrow$, Time $\downarrow$}\hfill
\figcap{\textbf{32.91}/\textbf{0.00043}, \textbf{4.5} ms}\hfill
\figcap{26.43/0.00218, 12.6 ms}\hfill
\figcap{31.24/0.00073, 15.7 ms}\hfill
\figcap{31.54/0.00066, 5.2 + 206 ms}\\
\caption{\label{fig:comp} {Comparison between reference, ours, ReSTIR~\cite{Lin2021} with 1 spp, ReSTIR with 1 spp and denoising, and MRPNN~\cite{Hu2023}.}}
\Description{}
\end{figure*} 

\begin{figure*}[ht!]
\centering
\newcommand{\figcap}[1]{\begin{minipage}{0.165\linewidth}\centering#1\end{minipage}}
\figcap{Reference}\hfill
\figcap{Ours}\hfill
\figcap{Reference}\hfill
\figcap{Ours}\hfill
\figcap{Reference}\hfill
\figcap{Ours} \vspace{0.25em}\\
\includegraphics[trim = 0 30 0 80, clip, width=0.165\linewidth]{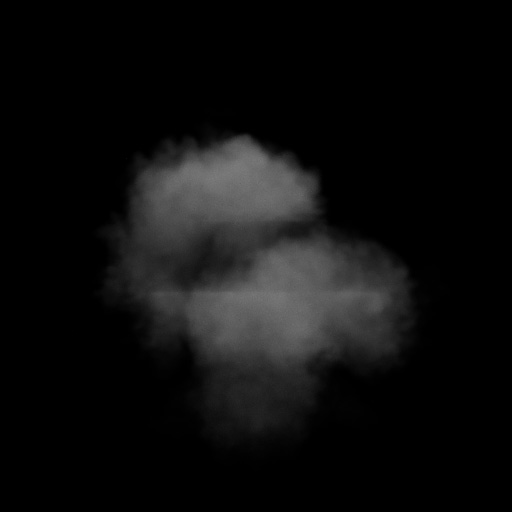}\hfill
\includegraphics[trim = 0 30 0 80, clip, width=0.165\linewidth]{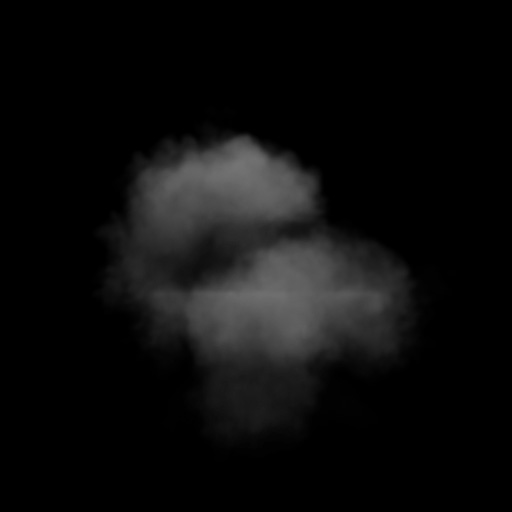}\hfill
\includegraphics[trim = 0 30 0 80, clip, width=0.165\linewidth]{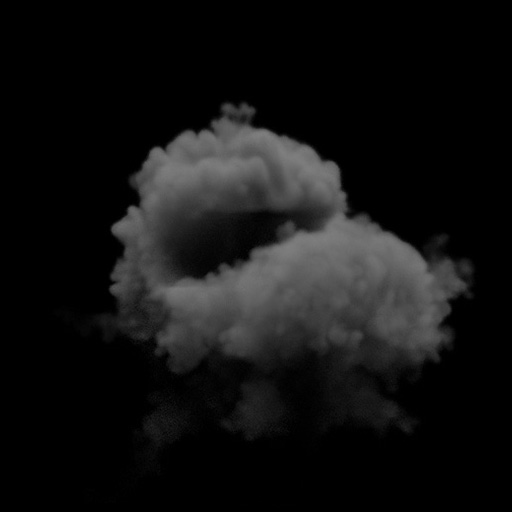}\hfill
\includegraphics[trim = 0 30 0 80, clip, width=0.165\linewidth]{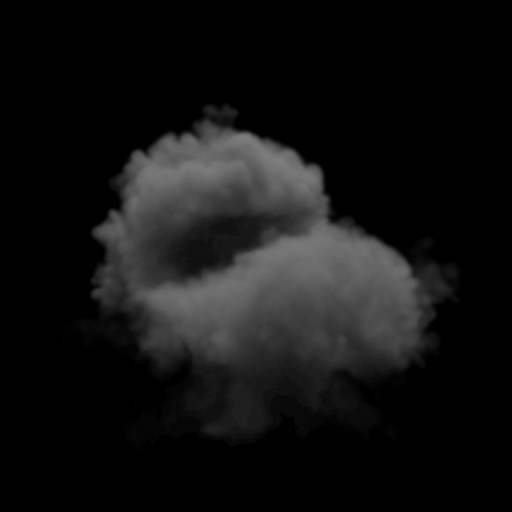}\hfill
\includegraphics[trim = 0 30 0 80, clip, width=0.165\linewidth]{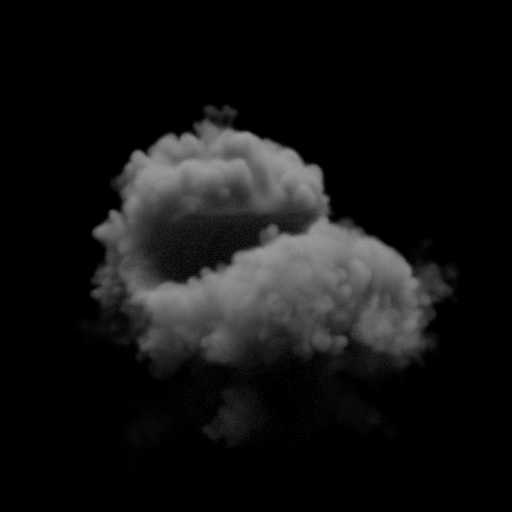}\hfill
\includegraphics[trim = 0 30 0 80, clip, width=0.165\linewidth]{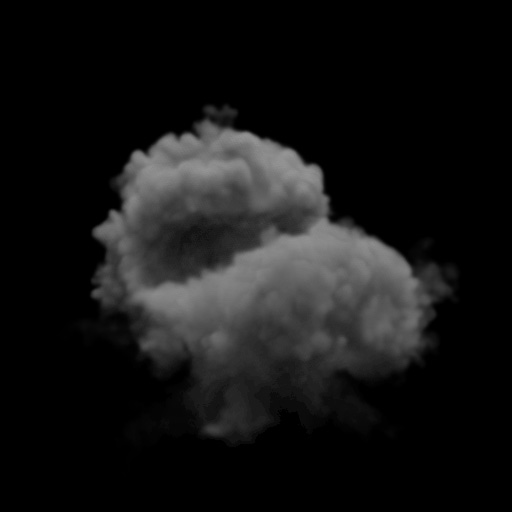}\\
\figcap{Density $\times 0.5$}\hfill
\figcap{32.39/0.00057}\hfill
\figcap{Density $\times 1.0$}\hfill
\figcap{40.71/0.00008}\hfill
\figcap{Density $\times 2.0$}\hfill
\figcap{35.89/0.00025} \vspace{-0.25em}
\caption{\label{fig:density} Ablation study on different densities. While PSNR decreases slightly due to the distribution shift, the results remain stable.}
\Description{}
\end{figure*} 

\begin{figure*}[ht!]
\centering
\newcommand{\figcap}[1]{\begin{minipage}{0.165\linewidth}\centering#1\end{minipage}}
\figcap{Reference}\hfill
\figcap{No channel adapter}\hfill
\figcap{No SRGB space}\hfill
\figcap{W/o flow loss}\hfill
\figcap{W/o perceptual loss}\hfill
\figcap{Our total} \vspace{0.25em}\\
\includegraphics[trim = 0 30 0 80, clip, width=0.165\linewidth]{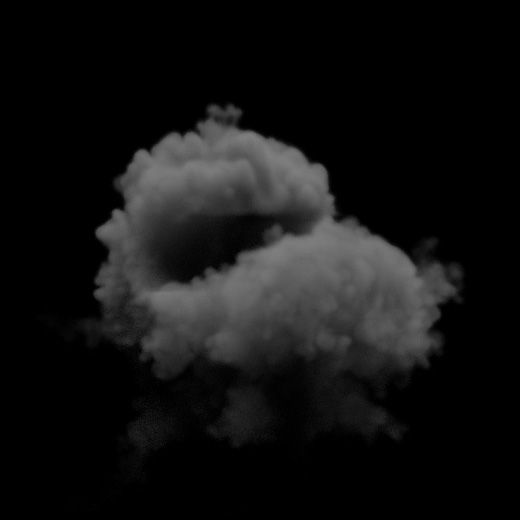}\hfill
\includegraphics[trim = 0 30 0 80, clip, width=0.165\linewidth]{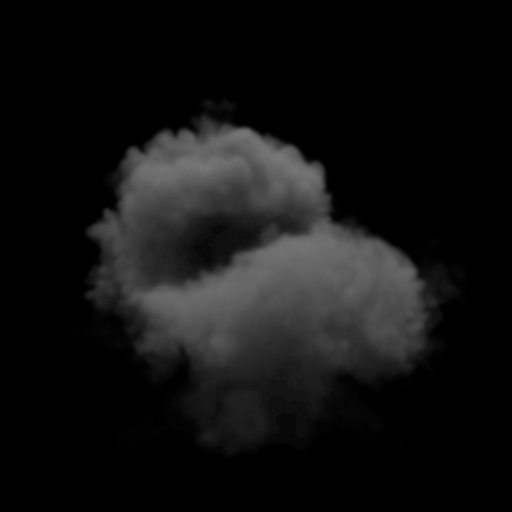}\hfill
\includegraphics[trim = 0 30 0 80, clip, width=0.165\linewidth]{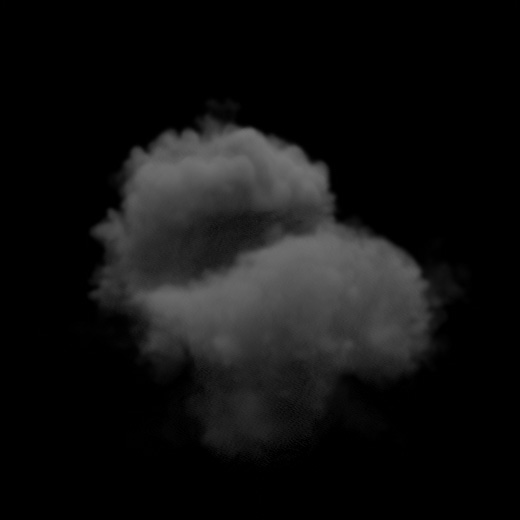}\hfill
\includegraphics[trim = 0 30 0 80, clip, width=0.165\linewidth]{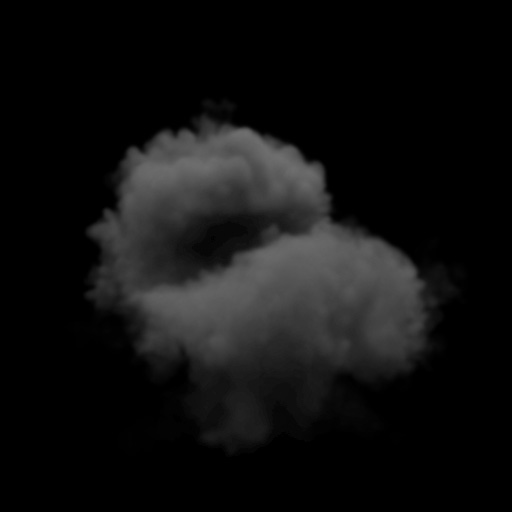}\hfill
\includegraphics[trim = 0 30 0 38, clip, width=0.165\linewidth]{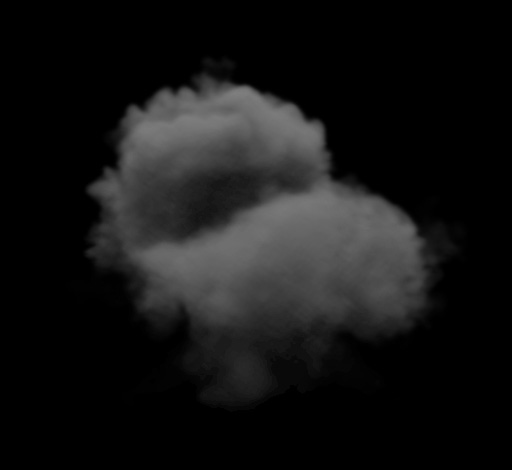}\hfill
\includegraphics[trim = 0 30 0 80, clip, width=0.165\linewidth]{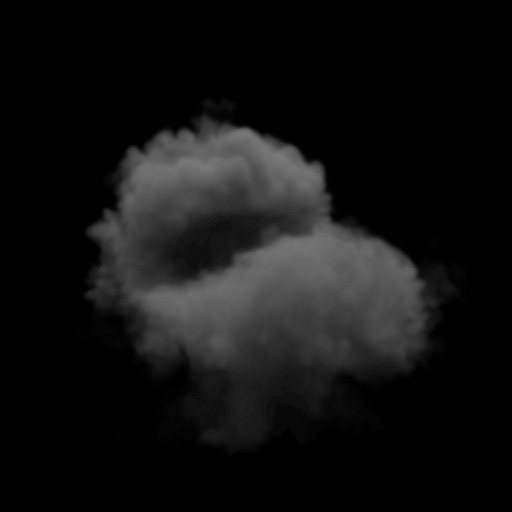} \\
\figcap{Avg./max/min PSNR $\uparrow$ }\hfill
\figcap{38.75/44.90/36.03}\hfill
\figcap{38.93/46.10/36.09}\hfill
\figcap{39.11/47.31/36.15}\hfill
\figcap{38.93/46.10/36.09}\hfill
\figcap{\textbf{40.85/48.78/37.93}}\\
\figcap{Avg./max/min MSE $\downarrow$ }\hfill
\figcap{\small 0.00019/0.00029/0.00002}\hfill
\figcap{\small 0.00017/0.00027/0.00002}\hfill
\figcap{\small 0.00016/0.00024/0.00001}\hfill
\figcap{\small 0.00017/0.00027/0.00003}\hfill
\figcap{\small \textbf{0.00009/0.00016/0.00001}}\vspace{-0.25em}
\caption{\label{fig:ablation} Ablation study on different losses. Each component proves essential for achieving high-quality results.}
\Description{}
\end{figure*}

\begin{figure*}[hb!]
\centering
\includegraphics[trim = 50 150 100 0, clip, width=0.199\linewidth]{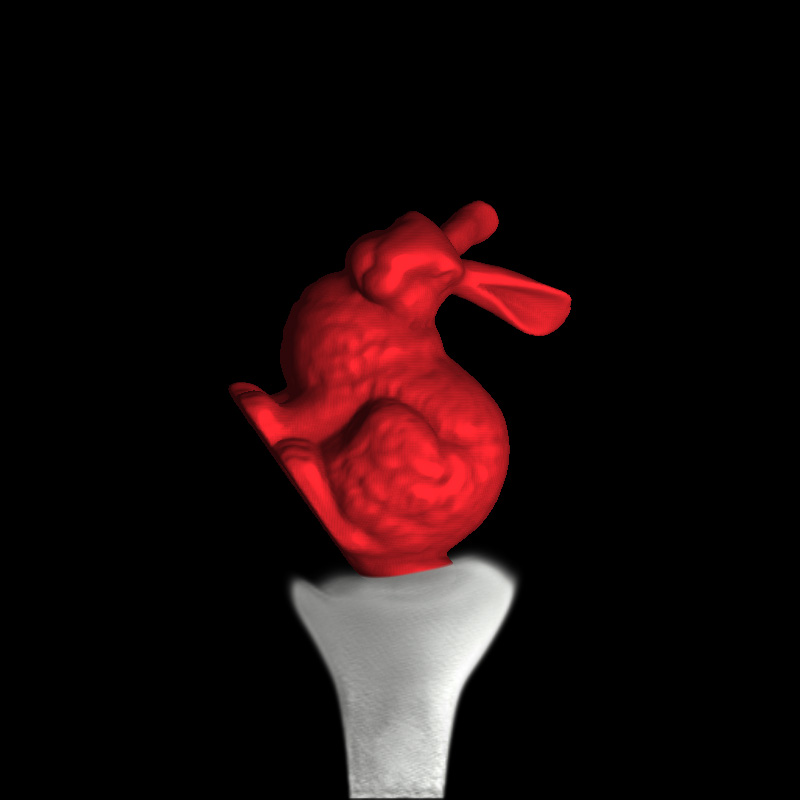}\hfill
\includegraphics[trim = 50 150 100 0, clip, width=0.199\linewidth]{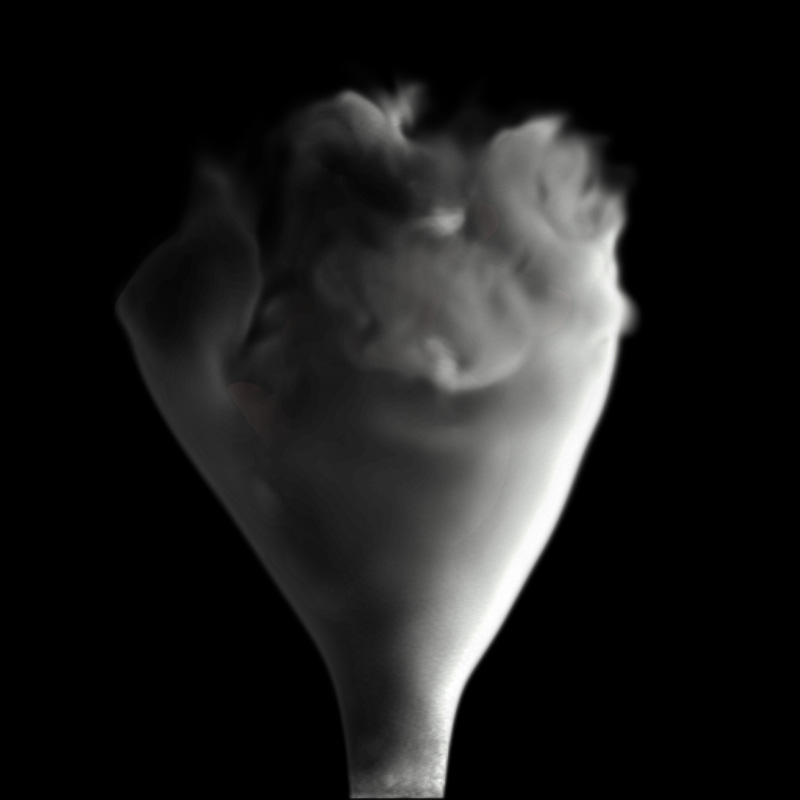}\hfill
\includegraphics[trim = 50 150 100 0, clip, width=0.199\linewidth]{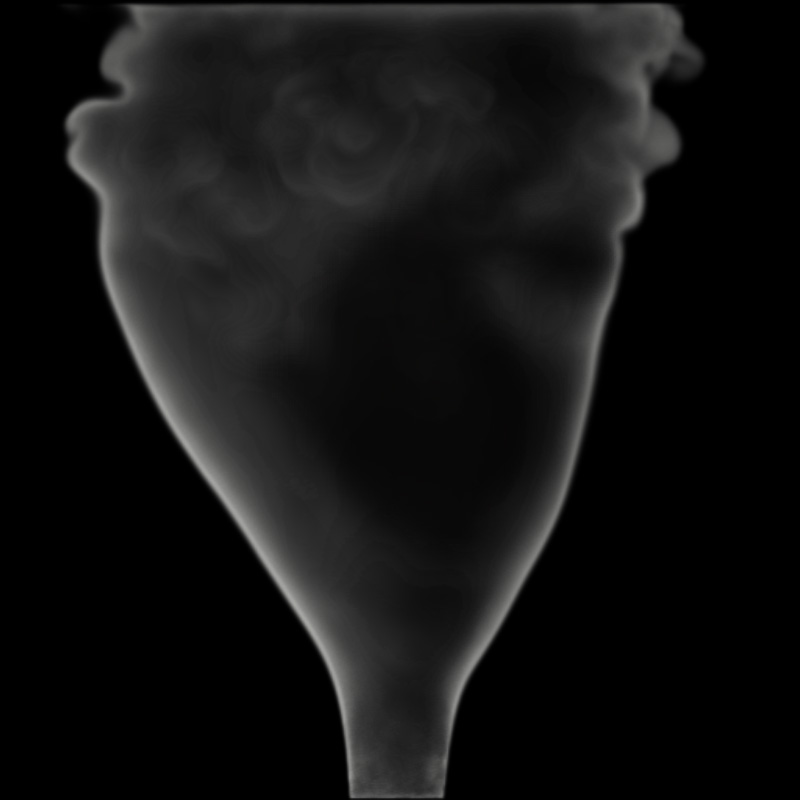}\hfill
\includegraphics[trim = 50 150 100 0, clip, width=0.199\linewidth]{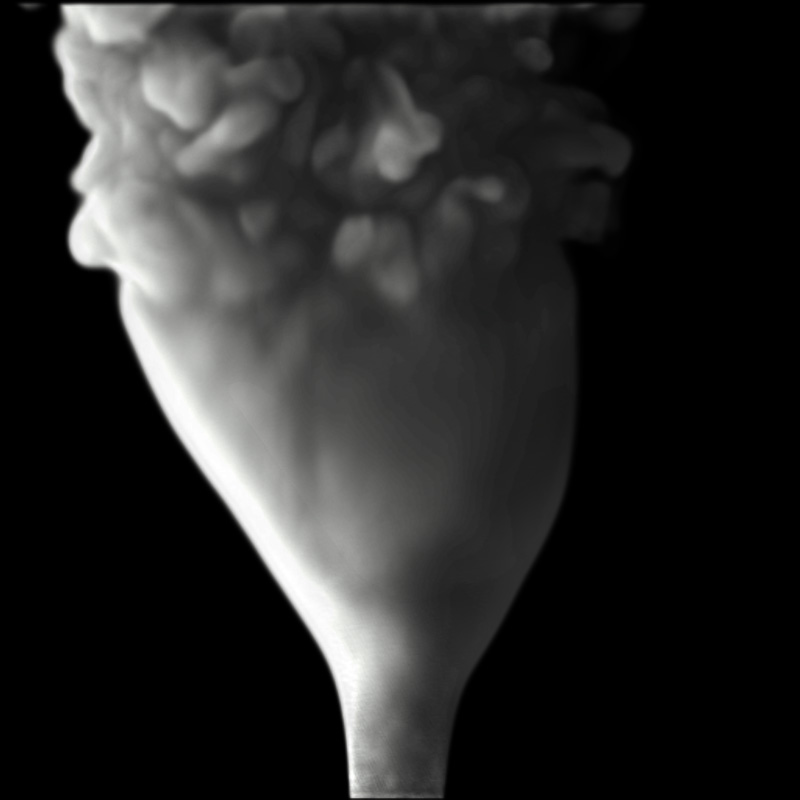}\hfill
\includegraphics[trim = 50 150 100 0, clip, width=0.199\linewidth]{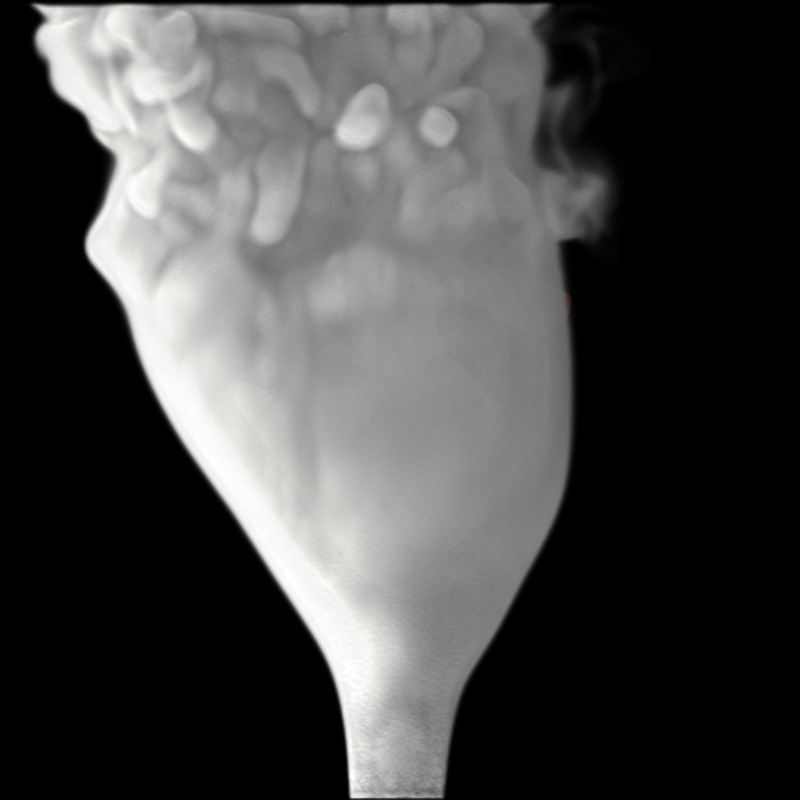}
\caption{\label{fig:bunny} Jet flow over a rigid bunny with a rotating light. The middle image demonstrates the bunny casting a shadow on the smoke from the backlit. }
\Description{}
\end{figure*}

\end{document}

%% file: intro.tex
\section{Introduction}
The world around us is filled with participating media, such as fog, smoke, clouds, and fire, which attenuate and scatter light in complex ways. Efficiently rendering volumetric media is a critical challenge for real-time applications, particularly in film and visual effects production, video games, and VR/AR. 
Conventional techniques, such as ray marching~\cite{Raab2008} and delta tracking~\cite{Novak2014}, are commonly used to simulate volumetric effects, relying on Monte Carlo (MC) path sampling to evaluate transmittance between two points in the scene for shadowing and fractional visibility. Although these techniques produce highly realistic results, they are computationally expensive because of complex lighting interactions within media, such as scattering and absorption, which require large numbers of samples for accurate evaluation in 3D. Recently, ReSTIR~\cite{Lin2021} was introduced to reuse paths to accelerate convergence, but it still relies on a denoiser to achieve real-time framerate with low samples per pixel.  

For real-time applications, smoke rendering is often achieved using textured sprites due to stringent computational constraints~\cite{unrealengine}. Recently, the \emph{six-way lightmaps} technique~\cite{Muller2023sixway} has emerged as a widely adopted method for approximating lighting effects on smoke, offering a favorable trade-off between realism and efficiency. This approach precomputes smoke transparency from a fixed view direction, along with light interactions from six orthogonal directions, using a sequence of pre-simulated 3D volumetric density fields. The results are stored in texture maps, which can then be efficiently sampled at runtime. During rendering, smoke is rendered as a camera-facing billboard and illuminated by interpolating between precomputed lightmaps for arbitrary light directions, enabling realistic visual effects such as internal shadowing and rim lighting. However, because both smoke dynamics and lighting are precomputed, this technique is inherently limited to static, pre-simulated sequences and fixed view directions (see~\autoref{fig:teaser}), restricting its applicability to interactive or dynamic environments.

To address the inherent limitations of the six-way lightmaps technique, we introduce a neural lightmaps method that learns light scattering from a 2D screen-space guiding map, in contrast to prior approaches that predict radiance directly in 3D space~\cite{Hu2023}. Given a 3D smoke density field generated by a physics-based simulator, our pipeline begins by ray-marching along the given view direction with a large sampling distance to generate a guiding map that approximates the smoke structure and silhouette. A U-Net with specialized channel adapters then uses this guiding map to predict the six-way lightmaps and the transparency map. These outputs can be seamlessly integrated into existing game engine rendering pipelines for final rendering. In addition, to approximate the shadow cast from objects onto the smoke, we construct a screen-space smoke depth map and compare it with the object shadow map during shading to determine whether each fragment lies within shadow. 
Our framework offers several advantages. First, the entire rendering pipeline completes in less than 4 ms, achieving an order-of-magnitude speedup over traditional Monte Carlo or learning-based 3D volume rendering methods for dynamic smokes. Second, our approach supports object–smoke interactions and dynamic camera movement, allowing real-time interactivity.
%
We evaluate our pipeline across a diverse set of scenes, including jet flows, explosions, and smoke interacting with rigid objects. We further demonstrate practical application scenarios by integrating our method into Unreal Engine for real-time rendering.


%% file: related.tex
\section{Related work}

\paragraph{Traditional Methods}
One common method for rendering participating media involves distributing light photons, rays, or beams within the media to estimate volumetric density~\cite{Jensen1998photon, Jarosz2011beam, jarosz11progressive}. This density is then combined with Monte Carlo path tracing to produce a unified solution~\cite{Krivanek2014Unifying}. Recent advancements~\cite{bitterli2017beyond, qin2015unbiased, Deng2019Photon} have improved convergence and achieved unbiased results. However, these methods require an additional light pass and extra storage to capture the light distribution. \emph{Virtual Point Lights (VPLs)}~\cite{keller1997instant} were extended for participating media by~\citet{arbree2008single}, enabling the approximation of subsurface scattering while significantly reducing the number of photons required. Building on this, \emph{Virtual Ray Lights (VRLs)}~\cite{Novak2012vrl}, clustered representations~\cite{Huo2016}, and hierarchical lighting grids~\cite{Yuksel2017} were introduced, further optimizing the computational and storage efficiency for volumetric rendering. 
Path tracing and Monte Carlo sampling offer a general framework for integrating the volume rendering equation, accommodating a wide variety of participating media scenarios. Numerous methods have been introduced for sampling free paths, including closed-form tracking~\cite{brown2003direct}, regular tracking~\cite{Brown1999, Hubert2006}, ray marching~\cite{Perlin1989, Raab2008, Munoz2014}, and null-collision techniques~\cite{Butcher1958, Georgiev2013, Kutz2017}. For a comprehensive overview of these techniques, we direct readers to surveys by~\citet{Wu2022survey} and~\citet{Novák2018survey}. Recently, \citet{Lin2021} proposed a low-noise, interactive volumetric path-tracing method by reusing scattering paths, significantly improving the efficiency of evaluating the volume rendering equation. However, despite such advancements, traditional volume rendering methods remain several orders of magnitude slower than the computational budgets typically allocated for real-time applications.

\paragraph{Learning-based Methods}
Several works have employed learning-based approaches to accelerate rendering tasks, such as predicting the radiance function~\cite{Kallweit2017}, the position of the exit point~\cite{Vicini2019}, and the contributions of all possible paths~\cite{Leonard2021}. Neural Radiance Fields (NeRF)~\cite{Mildenhall2021NeRF} represent scenes as neural radiance fields and render with ray marching, accelerated by~\citet{Muller2021}. Recently, \citet{Hu2023} presented a lightweight feature-fusion neural network capable of rendering high-order scattered radiance from participating media in real time. However, these methods primarily focus on accelerating Monte Carlo rendering in 3D world space, where the computational complexity of 3D operations inherently constrains the acceleration ratio. More importantly, it relies on heavy precomputation, which prevents it from being used in dynamic scenes. Instead, our method brings neural rendering one step closer to real-time by learning lightmaps based on screen-space input.

\paragraph{Relighting}
Another line of work related to our method involves using lightweight neural networks combined with physically based priors for relighting. These approaches typically take monocular images as input and generate relighted outputs. Previous methods in this area derive priors from various domains such as outdoor scenes~\cite{wu2017interactive, liu2020learning}, portraits~\cite{pandey2021total, ranjan2023facelit}, and human bodies~\cite{kanamori2019relighting, ji2022geometry}. With advancements in generative models, more sophisticated techniques, such as diffusion-based networks, e.g., Dilightnet~\cite{zeng2024dilightnet}, Neural Gaffer~\cite{jin2024neural}, and sought-after IC-light~\cite{zhang2025scaling}, have made significant progress in producing relightable images. However, these methods focus on diffuse or specular materials and surface rendering, with limited exploration of volumetric rendering.






%% file: background.tex

\section{Traditional Six-way Lightmaps}
In this section, we provide a brief review of the volume rendering equation that governs smoke appearance, as well as the six-way lightmaps technique for real-time smoke rendering. 

The volume rendering equation represents incident radiance $L(\xx,\om)$ at the point $\xx$ from the outgoing direction $\om$, which is derived by integrating the outgoing radiance through volumetric media and accounting for the surface or light $L_o$ behind it multiplied by the transmittance $T(\xx \leftrightarrow \zz)$ along the ray, formalized as:
\begin{align}
\label{eq:RE}
L(\xx, \om) = & \int_{0}^{z} T(\xx \leftrightarrow \yy) \bigl[ \sigma_a(\yy) L_e(\yy, \om) + \sigma_s(\yy) L_s(\yy, \om) \bigr] dy \\ 
& + \; T(\xx \leftrightarrow \zz) L_o(\zz, \om)\;, \nonumber 
\end{align}
where $\yy = \xx - y \om$ is a point along the direction $\om$ towards $\xx$, and $\sigma_a(\xx)$ and $\sigma_s(\xx)$ denote the absorption and scattering coefficients, respectively. The emitted radiance is represented by $L_e$, while the in-scattered radiance term:
\begin{align} \label{eq:scatter}
L_s(\yy, \om) = \int_S \rho(\yy,\om,{\om}') L(\yy,{\om}') d{\om}'
\end{align} 
integrates incident light over the sphere $S$ of all directions, modulated by $\rho(\yy,\om,{\om}')$, the media phase function. And $L(\yy,{\om}')$ denotes the incoming radiance at the point $\yy$ from the direction ${\om}'$, comprising both unscattered (direct) light and light that has experienced an arbitrary number of scattering events prior to reaching $\yy$. If only one single directional light source is considered, the in-scattered radiance term can be further simplified to a function of the light direction $\om_l$ as $L_s(\yy, \om, \om_l)$. 
%
The transmittance function
\begin{align}
T(\xx \leftrightarrow \yy) = e^{-\int_{0}^{y} \sigma_t(\xx - s \om) ds }
\end{align} 
represents visibility between points $\xx$ and $\yy$, where the extinction coefficient $\sigma_t(\xx) = \sigma_a(\xx) + \sigma_s(\xx)$, and $L_o(\zz, \om)$ denotes the background radiance along $\om$.
The nested integrals in~\autoref{eq:RE} are computationally expensive to approximate, making Monte Carlo sampling–based methods impractical for real-time rendering.

\begin{figure}[t!]
\centering
\includegraphics[width=\columnwidth]{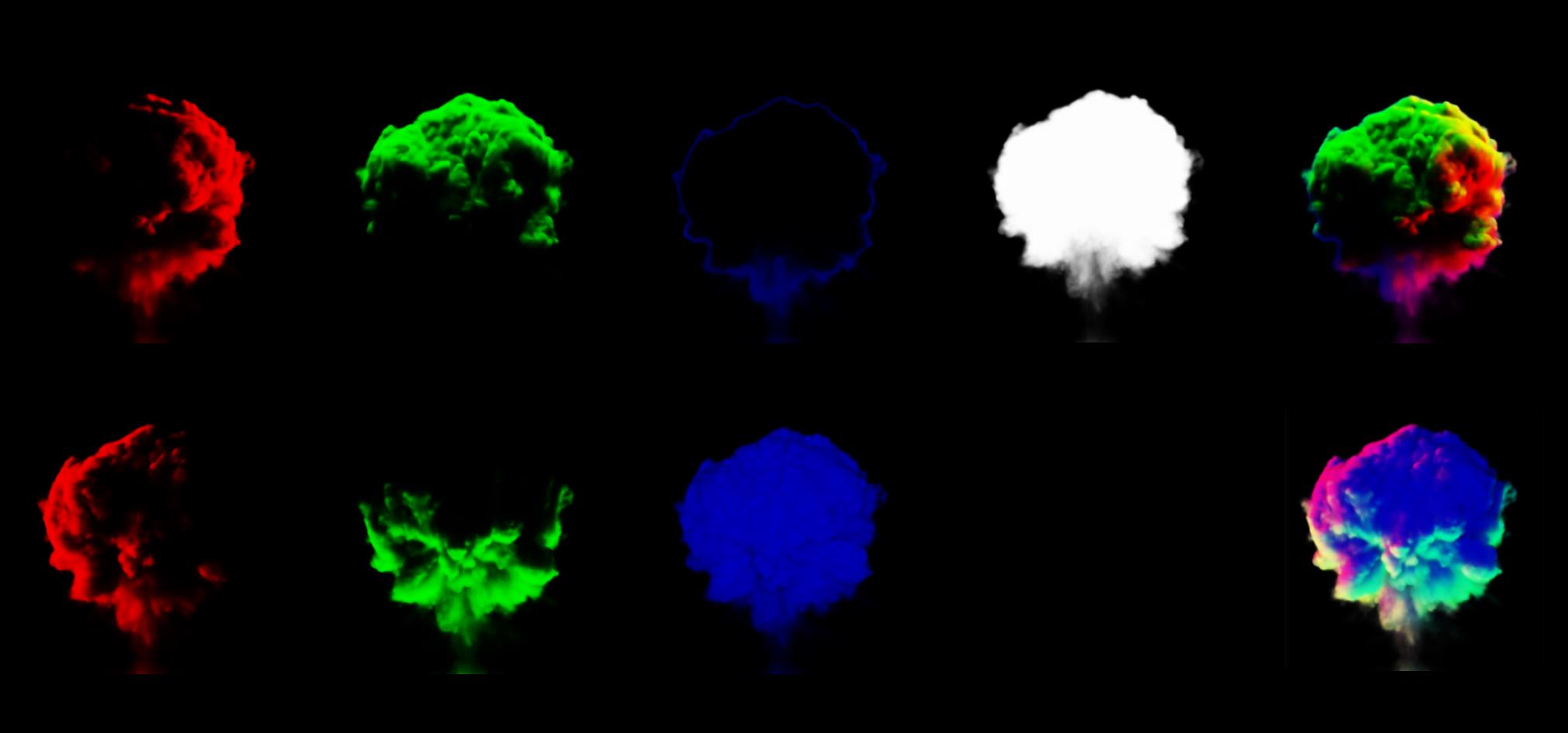}
\put(-226,104){\small \textcolor{red}{Red}}
\put(-182,104){\small \textcolor{green}{Green}}
\put(-130,104){\small \textcolor{blue}{Blue}}
\put(-83, 104){\small \textcolor{gray}{Alpha}}
\put(-35, 104){\small \textcolor{red}{R}\textcolor{green}{G}\textcolor{blue}{B}\textcolor{gray}{A}}
\put(-200, 78){\small \textcolor{white}{+}}
\put(-150, 78){\small \textcolor{white}{+}}
\put(-100, 78){\small \textcolor{white}{+}}
\put(-50,  78){\small \textcolor{white}{=}}
\put(-200, 30){\small \textcolor{white}{+}}
\put(-150, 30){\small \textcolor{white}{+}}
\put(-100, 30){\small \textcolor{white}{+}}
\put(-50,  30){\small \textcolor{white}{=}}
\put(-228, 55){\footnotesize \textcolor{red}{Right}}
\put(-177, 55){\footnotesize \textcolor{green}{Top}}
\put(-129, 55){\footnotesize \textcolor{blue}{Back}}
\put(-92,  55){\footnotesize \textcolor{gray}{Transparency}}
\put(-38,  55){\footnotesize \textcolor{white}{Lightmap 1}}
\put(-226, 3){\footnotesize \textcolor{red}{Left}}
\put(-180, 3){\footnotesize \textcolor{green}{Bottom}}
\put(-130, 3){\footnotesize \textcolor{blue}{Front}}
\put(-86,  33){\footnotesize \textcolor{gray}{Emissive}}
\put(-88,  26){\footnotesize \textcolor{gray}{(Optional)}}
\put(-38,  3){\footnotesize \textcolor{white}{Lightmap 2}}
\caption{\label{fig:sixway_example} An example set of lightmaps packed in two RGBA textures. Of the 8 available channels, 6 channels store the six-way scattering lightmaps along axis-aligned directions. Additionally, the alpha channel for the first texture contains the transparency $T(\xx \leftrightarrow \zz)$, while the alpha channel of the second texture is an optional emissive component.
}\vspace{-1em}
\Description{}
\end{figure}

\emph{Six-way lightmaps} technique~\cite{Muller2023sixway,Mai2023flipbook} provides an efficient and visually compelling approach for rendering volumetric effects such as smoke in real time, particularly in AAA game production.
In particular, this technique assumes that the smoke is always observed from a fixed view direction $\om$, enabling it to be rendered as a camera-facing billboard and for transparency and scattering to be precomputed.
Under this assumption, the volumetric rendering equation~(\autoref{eq:RE}) can be rewritten in following form:
\begin{align}
L(\xx)  = \; & \underbrace{\int_{0}^{z} T(\xx \leftrightarrow \yy) \sigma_a(\yy) L_e(\yy) dy }_{\text{emissive}}  \\
& + \underbrace{\int_{0}^{z} T(\xx \leftrightarrow \yy) \sigma_s(\yy) L_s(\yy, \om_l) dy }_{\text{scattering}} 
 + \underbrace{T(\xx \leftrightarrow \zz)}_{\text{transparency}} L_o(\zz)\;, \nonumber 
\end{align}
where transparency and emissive terms can be precomputed and stored as texture channels, since they are independent of the light direction $\om_l$. 
In contrast, the scattering term, defined in~\autoref{eq:scatter}, depends on the light direction $\om_l$ and is approximated by interpolating between six pre-rendered scattering lightmaps from the directions $\pm X$, $\pm Y$, $\pm Z$, denoted as $\mathcal{L} = \{L_x^+, L_x^-, L_y^+, L_y^-, L_z^+, L_z^-\}$. These lightmaps, together with transparency and emissive color, are packed into the RGBA channels of a texture, as illustrated in~\autoref{fig:sixway_example}. 

During rendering, for an arbitrary light direction $\om_l$, the total scattered radiance at position $\xx$ is approximated by a directionally weighted interpolation:
\begin{align}
L_\text{scattering}(\xx, \om_l) &= \int_{0}^{z} T(\xx \leftrightarrow \yy) \sigma_s(\yy) L_s(\yy, \om_l) dy \\ &\approx \sum_{p \in \{x, y, z\}} |\om_{l,p}|L_p^{\text{sign}(\om_{l,p})}(\xx) \;, \nonumber
\end{align}
where $\om_{l,p}$ denotes the component of $\om_l$ along axis $p$. This technique naturally supports an arbitrary number of directional lights. The total scattered radiance is obtained by summing the contributions from all light $L_\text{scattering}(\xx, \om_l)$, each weighted by its light intensity. It is also easy to support multiple light sources by simply summing all light contributions.


Six-way lightmaps typically require two $512\sqrt{K}\times 512\sqrt{K}$ RGB(A) textures to store $K$ frames of pre-simulated smoke dynamics, with each frame rendered at $512 \times 512$ resolution. At runtime, the shading point's relative light direction is calculated and used to blend between these lightmaps dynamically. 
%
%
Although the results differ from ground truth 3D volume rendering due to interpolations on lightmaps, its real-time performance makes it ideal for background effects like dust storms, tornadoes, and volumetric clouds with dynamic lighting. 
However, since the smoke sequence and lighting are baked into textures, smoke dynamics, camera settings, and lighting are pre-simulated and pre-rendered from a fixed view direction, limiting the interactivity of this technique, as summarized in~\autoref{tab:compare_methods}. 

\begin{table}[h!] \vspace{-0.5em}
\caption{Comparison between traditional and our neural six-way lightmaps. Dyn. denotes dynamic. (No$^\star$: Traditional method renders a billboard under any view, which introduces large artifact as shown in~\autoref{fig:teaser})} \vspace{-1em}
\label{tab:compare_methods}
\newcommand{\bad}{\cellcolor{red!20}}
\newcommand{\ok}{\cellcolor{yellow!20}}
\newcommand{\good}{\cellcolor{green!20}}
\newcommand{\non}{\cellcolor{white!20}}
%
\newcommand{\checked}{\cellcolor{green}}
%
\scalebox{0.8}{
\begin{tabular}{l|ccccc}
\toprule
\non \textbf{Method}&\non\textbf{Dyn. light}&\non\textbf{Dyn. smoke}&\non\textbf{Dyn. view}&\non\textbf{Obstacle}&\non\textbf{Speed}\\
\midrule
\non Traditional & \good Yes & \bad No & \bad No$^\star$ & \bad No & \good Very fast  \\
\non Ours neural & \good Yes & \good Yes & \good Yes & \good Yes & \ok Fast \\
\bottomrule
\end{tabular}
} \vspace{-1.5em}
\end{table}

%% file: method.tex
\section{Neural Six-way Lightmaps Prediction}

In this section, we introduce our neural six-way lightmaps technique, which enables real-time performance while allowing dynamic camera direction and arbitrary user interactivity. 
%

\begin{figure*}[ht!]
\centering
\includegraphics[width=\linewidth]{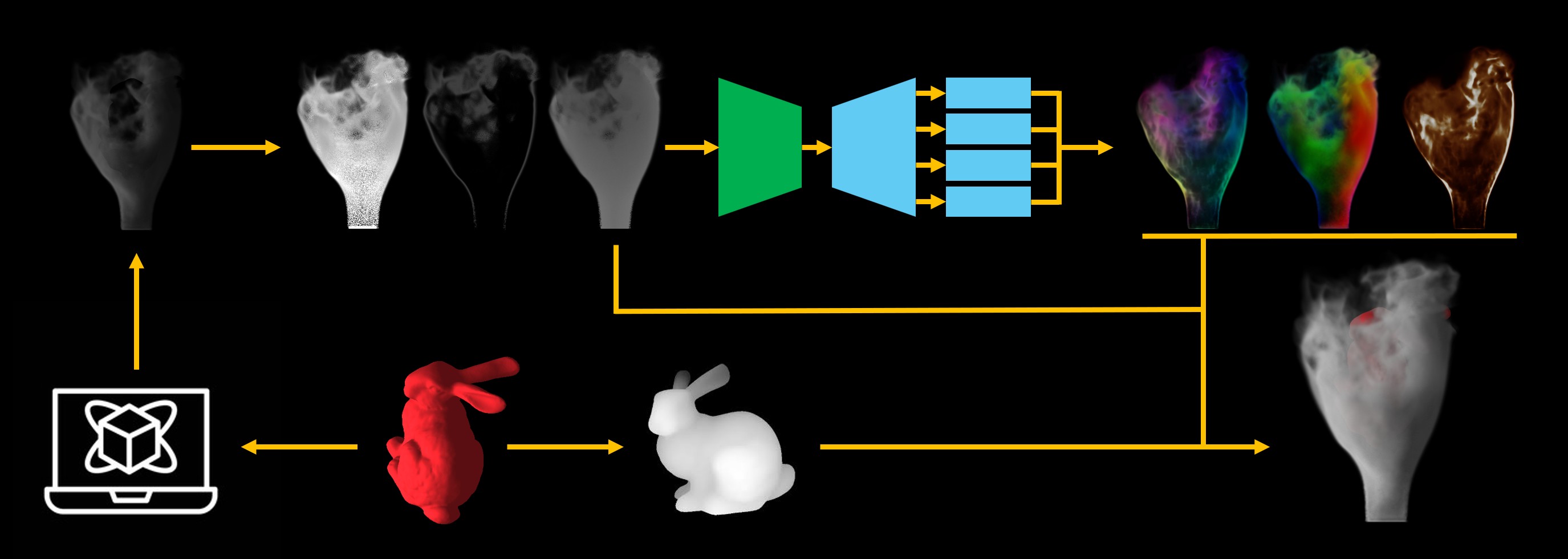}
\put(-488,4){\footnotesize \textcolor{white}{Fluid simulator}}
\put(-380,4){\footnotesize \textcolor{white}{Obstacle}}
\put(-290,4){\footnotesize \textcolor{white}{Shadow map}}
\put( -75,4){\footnotesize \textcolor{white}{Final result}}
\put(-487,172){\footnotesize \textcolor{white}{Density grid}}
\put(-408,172){\footnotesize \textcolor{white}{$\Tilde{L}_{\text{scattering}}$}}
\put(-355,172){\footnotesize \textcolor{white}{$T$}}
\put(-315,172){\footnotesize \textcolor{white}{$D$}}
\put(-275,132){\footnotesize \textcolor{black}{Encoder}}
\put(-238,132){\footnotesize \textcolor{black}{Decoder}}
\put(-197,149){\footnotesize \textcolor{black}{F \& B}}
\put(-197,137){\footnotesize \textcolor{black}{L \& R}}
\put(-197,125){\footnotesize \textcolor{black}{U \& D}}
\put(-197,114){\footnotesize \textcolor{black}{A \& E}}
\put(-213,102){\footnotesize \textcolor{white}{Channel adapters}}
\put(-131,172){\footnotesize \textcolor{white}{Six-way lightmaps}}
\put(-57,172){\footnotesize \textcolor{white}{Emissive (optional)}}
\put(-415,28){\footnotesize \textcolor{white}{(a)}}
\put(-477,78){\footnotesize \textcolor{white}{(b)}}
\put(-437,125){\footnotesize \textcolor{white}{(c)}}
\put(-290,125){\footnotesize \textcolor{white}{(d)}}
\put(-331,28){\footnotesize \textcolor{white}{(e)}}
\put(-122,28){\footnotesize \textcolor{white}{(f)}}
\caption{\label{fig:pipeline} Our pipeline: the physically based fluid simulator takes the obstacle as input (a) to produce the density field (b). (c) A ray marching with a large sample step extracts the guiding map with three channels, in-scattered radiance $\Tilde{L}_{\text{scattering}}$, transparency $T$, and depth $D$. (d) Our neural lightmaps generator contains a modified UNet that first extracts channel-shared features from the input, which are then processed by four dedicated channel adapters with outputs for front and back, left and right, up and down, and transparency and emissive, respectively. The shadow map of obstacle (e) is generated from the light and combined with predicted lightmaps to generate  the final rendering result (f).}  \vspace{-0.5em}
\Description{}
\end{figure*}

\subsection{Overview}
Unlike prior learning-based volumetric rendering approaches that directly accelerate the evaluation of Radiative Transfer Equation in 3D, our method draws inspiration from traditional six-way lightmaps and leverages screen-space information to predict integrated light scattering from six orthogonal directions. This design dramatically reduces network complexity, simplifies training, and yields orders-of-magnitude performance improvements.
As illustrated in \autoref{fig:pipeline}, our system begins with a real-time 3D GPU fluid solver that supports dynamic smoke–obstacle interactions. To approximate volumetric scattering efficiently, we generate a screen-space guiding map and a depth map by ray-marching the density field along the camera direction with a large step size under a surrogate lighting setup. The resulting 2D guiding map is fed into a compact neural network that predicts the six directional lightmaps. These lightmaps, combined with the obstacle's shadow map, enable high-quality smoke illumination in a few milliseconds, supporting dynamic smoke, arbitrary camera motion, and changing light conditions.

\subsection{Guiding Map via Ray-marching}
Estimating the rendering equation~\autoref{eq:RE} within the limited time budget of real-time applications is computationally challenging. Modern high-quality, physically based volumetric renderer relies on stochastic procedures to determine the distance and direction to the next sample while simultaneously estimating the transmittance along the ray. Although accurate, this approach requires significant computational overhead, making it challenging to achieve real-time performance.

Our goal is to efficiently generate a coarse approximation of the smoke’s structure and silhouette, which serves as a guide for our network to predict the corresponding lightmaps. To this end, we illuminate the smoke using three carefully selected light directions. The first light is aligned with the reversed viewing direction $\om$, as the smoke’s appearance from the viewing direction is most critical for its on-screen representation. To further emphasize the silhouette and enhance structural perception, we introduce two additional side lights from directions $\om \times \zz$ and $-\om \times \zz$, where $\zz$ denotes the z-axis in world-space. Hence, we adopt an aggressive approximation of the in-scattered radiance $\Tilde{L}_{\text{scattering}}(\xx, \om)$ as :
\begin{align} \label{eq:approx}
\Tilde{L}_{\text{scattering}}(\xx, \om) = \int_{0}^{z} T(\xx \leftrightarrow \yy) \sigma_s(\xx) L_s (\yy) dy \;, 
\end{align}
where $L_s (\yy) = \sum_{{\om}'\in\{\text{front},~\text{top},~\text{bottm} \}}  T(\yy \leftrightarrow \zz_{{\om}'}) L_{{\om}'}$ and front, top, and bottom denotes $\om$, $\om \times \zz$, and $-\om \times \zz$, respectively. In practice, to further accelerate the computation, we use ray-marching with a large step size to obtain a rough approximation of~\autoref{eq:approx}, as described in~\autoref{alg:mainSim}.

\RestyleAlgo{ruled}
\begin{algorithm} [t!]
\caption{Ray-marching for Guiding Map $\Tilde{L}_{\text{scattering}}$}
\label{alg:mainSim}
\DontPrintSemicolon
\newcommand\mycommfont[1]{\footnotesize\ttfamily\textcolor[RGB]{0 128 0}{#1}}
\SetCommentSty{mycommfont}
\KwIn{Time step size $h$, number of steps $N$ }
\KwOut{$\Tilde{L}_{\text{scattering}}$, $D$, and $T_N$}
\For{each view ray from $\xx$ along direction $\om$}{
    $\Tilde{L}_{\text{scattering}}\gets0$, $c_0\gets1$, $D\gets0$  \\
    $\xx_0 \gets \xx_0 + \delta$  \tcp*[l]{$\delta$: random jitter} 
    \For{$n \gets 1$ \textbf{to} $N$}{
        $\xx_n \gets\xx_0- n h \om$, \, $\sigma_n\gets\sigma_s(\xx_n)$\\
        \If{$D = 0$ \textbf{and} $\sigma_n > \tau$}{
            $D \gets n h$ \\
        }
        $T_n\gets T_{n-1}e^{-\sigma_nh}$, \, $A_n\gets T_{n-1}\sigma_s$ \\
                 \tcp*[l]{ $A^{\bullet} L^\bullet P^\bullet$ computed via ray-marching along direction $\bullet$ }  \tcp*[l]{with large step size}
        $L_n\gets T^{\text{front}} L^{\text{front}} P(\om, \text{front}) + T^{\text{top}} L^{\text{top}} P(\om, \text{top}) + T^{\text{bottom}} L^{\text{bottom}} P(\om, \text{bottom})$ \\
        $\Tilde{L}_{\text{scattering}}\gets \Tilde{L}_{\text{scattering}}+ A_n L_n$
    }
} 
\end{algorithm} 

We adopt \autoref{alg:mainSim} as a guiding map generator, using a ray marching with a large step size of $h=10 \Delta x$ in smoke data $\sigma_s$ in a 3D texture, and $\Delta x$ is the voxel width. Our output guiding maps $\mathcal{G}$ involve three channels, including the approximated in-scattered radiance $\Tilde{L}_{\text{scattering}}$, transparency $T$, and depth $D$ of the smoke surface to the camera. The rendering result, as illustrated in~\autoref{fig:pipeline}, shows that while the guiding map lacks fine-scattering details and exhibits blocky artifacts due to its coarse step size, it effectively captures the essential smoke structure and silhouette. This approximation is sufficient to guide a neural network in generating six-way lightmaps. We add a small jitter distance $\delta$ to each marching step to avoid the pattern artifacts that appear when using large ray marching steps.



\subsection{Channel-aware Neural Lightmaps Generator}

We denote by $\mathcal{G}^t \in \mathbb{R}^{H \times W \times 3}$ the guiding map generated via ray marching in the $t$-th frame. Our objective is to design a neural network that translates the guiding map $\mathcal{G}^t$ into the final rendering results that closely approximate those produced by ground-truth Monte Carlo integration. A straightforward approach is to feed both the guiding map and the light direction into the network and directly predict the final radiance. However, this design suffers from two major drawbacks. First, it requires sampling over light directions during training, significantly increasing both data generation cost and training complexity. Second, in scenes that are illuminated by multiple light sources, a common scenario in modern game environments, the inference time scales linearly with the number of light sources, making real-time performance impractical.

Hence, we adopt the technique inspired by current game rendering pipelines and predict six-way lightmaps $\mathcal{L}^t \in \mathbb{R}^{H \times W \times 6}$, where each channel corresponds to one of the six canonical directions and $t$ denotes the current frame. These lightmaps enable smoke shading to be computed efficiently via directional interpolation. The network must satisfy several key requirements: (1) perform inference in real time to meet the stringent performance demands of video games, (2) produce temporally coherent outputs across consecutive frames, and (3) generalize effectively to unseen smoke density distributions $\sigma_s$ generated by the real-time smoke simulator. While prior work, such as~\cite{chu2020learning}, has explored GAN-based approaches with spatial–temporal discriminators trained on 3D density data, their inference costs remain prohibitively high for real-time applications. To achieve efficiency, we instead rely solely on the 2D screen-space guiding map $\mathcal{G}^t$ as network input.


As illustrated in~\autoref{fig:pipeline}, we parameterize our lightmap generator using a modified UNet~\cite{ronneberger2015u} architecture with channel adapters, which serve as the generator translating $\mathcal{G}^t$ to $\mathcal{L}^t$. The UNet encoder extracts a bottleneck feature $f^t \triangleq \text{Enc}(\mathcal{G}^t)$, which the decoder subsequently transforms into the output lightmaps, $\mathcal{L}^t \triangleq \text{Dec}(f^t)$.
Observing that the output channels exhibit substantial directional variation yet share partial structural similarities, we group them into four categories: front-back lightmaps, left–right lightmaps, up–down lightmaps, and a group comprising transmittance and emissive color.

To better capture group-specific characteristics, we augment the UNet decoder with four dedicated channel adapters: the first three are directional groups, and the last is for alpha-emissive, as shown in~\autoref{fig:pipeline}.
These adapters, implemented as NAFBlocks~\cite{chen2022simple}, disentangle inter-channel dependencies, enabling the network to capture group-specific illumination properties and thereby improving both reconstruction fidelity and robustness.



\subsection{Training Procedure}

To train our network as a generalizable image translator, we construct a dataset of scattering coefficient trajectories ${\sigma^t_{s}}$ generated using an in-house fluid simulator. For each frame, we employ both ray marching and Monte Carlo (MC) integration to produce the ground-truth tuple $\langle \sigma^t_{s}, \mathcal{G}^t, \mathcal{L}^t \rangle$. During training, we optimize a composite objective designed to promote generalization, temporal coherence, and visual similarity between the predicted and reference six-way lightmaps, $\hat{\mathcal{L}}^t$ and $\mathcal{L}^t$, respectively.

Our objective consists of three loss components. The first is a pixel-wise reconstruction loss, formulated as the mean squared error (MSE) between the predicted and ground-truth lightmaps in sRGB space to better capture details in dark regions:
\begin{align}
l_\text{MSE} = |\hat{\mathcal{L}}^t - \mathcal{L}^t|_2^2 \;.
\end{align}
The second component is a perceptual loss~\cite{johnson2016perceptual}, which leverages a pretrained VGG network $\phi(\cdot)$ to extract hierarchical feature representations. We compute the $\ell_1$ distance between the feature maps of the prediction and the ground truth to enforce perceptual similarity:
\begin{align}
l_\text{perc} = |\phi(\hat{\mathcal{L}}^t) - \phi(\mathcal{L}^t)|_1 \;.
\end{align}
The third term enforces temporal consistency in the flow space using a pretrained FlowNet model $\text{O}(\cdot)$. Given paired frames $\langle \mathcal{L}^{t-1}, \mathcal{L}^t \rangle$ and $\langle \hat{\mathcal{L}}^{t-1}, \hat{\mathcal{L}}^t \rangle$, we compute their corresponding optical flow fields $v^t = \text{O}(\mathcal{L}^{t-1}, \mathcal{L}^t)$ and $\hat{v}^t = \text{O}(\hat{\mathcal{L}}^{t-1}, \hat{\mathcal{L}}^t)$. The flow loss is then defined as the $\ell_1$ difference between the predicted and reference flow fields:
\begin{align}
l_\text{flow} = |\hat{v}^t - v^t|_1 \;.
\end{align}
The overall training objective combines the three losses as a weighted sum:
\begin{align}
l = \lambda_\text{MSE}l_\text{MSE} + \lambda_\text{perc}l_\text{perc} + \lambda_\text{flow}l_\text{flow} \;,
\end{align}
where $\lambda_\text{MSE}$, $\lambda_\text{perc}$, and $\lambda_\text{flow}$ are scalar weights that balance the contributions of the individual loss terms.

\subsection{Obstacle Shadowing}
Unlike traditional six-way lightmaps, our method supports shadows cast by scene geometry onto participating media. Computing such shadows accurately typically requires tracing secondary rays through both the medium and surrounding obstacles, which is prohibitively expensive. Alternatively, encoding explicit geometry as part of the network input would increase training cost and data complexity, as meshes must be embedded into latent representations while preserving generalization.
To balance efficiency and fidelity, we employ a depth-based screen-space shadow approximation.
In particular, during the generation of the guiding map, we already obtain the depth of the smoke in screen space, which forms a smoke outer shell.
%
During rendering, each pixel's smoke ``surface'' depth is transformed into the light view–projection space and compared against a shadow map rendered from the light. By comparing the smoke shell depth with the obstacle shadow depth, we efficiently estimate per-pixel visibility without performing a second ray-marching pass for shadow evaluation.



%% file: result.tex

\section{Experiments}
We conducted all experiments on a workstation equipped with an AMD Ryzen Threadripper 3970X 32-core CPU, 256 GB of memory, and a GPU with 16,384 cores and 24 GB of memory. Following common game-engine practice, all evaluations use a $512\times512$ lightmap resolution to balance texture memory and performance. All presented results are generated from the same network.

\paragraph{Smoke Simulator}
To support interactive smoke dynamics with efficient rigid coupling, we employ a 3D physically based fluid solver based on the lattice Boltzmann method (LBM)~\cite{li2023high}. The LBM-generated velocity field advects the density field, which is streamed directly into the guiding map generation. This formulation inherently supports collisions with arbitrary scene geometry, making it well-suited to dynamic and interactive scenarios.

\paragraph{Data Generation}
We generate our smoke dataset by simulating 14 dynamic smoke sequences with an LBM solver, placing cylindrical obstacles at different locations to produce smoke with varying shapes. For each sequence, we render only the smoke while masking out the obstacles. High-fidelity offline volume rendering is used to generate six-way light maps and an alpha channel for 200 frames per sequence, which takes approximately 2 hours. Each sequence is rendered from 9 viewpoints by rotating the camera $90^\circ$ in $10^\circ$ increments, yielding a total of 126 smoke sequences and 25,200 images. We generate the reference images using the Houdini Karma renderer with volume path tracing. We configure the renderer with a single bounce, image resolution of $512\times512$ pixels, Henyey-Greenstein phase function with $g=0$, and 512 samples per pixel. The rendering time is approximately 20 seconds per image.

\paragraph{Training} We train our network using Adam optimizer with a learning rate of $0.001$. The batch size is set to $12$, and the training is conducted for $200$ epochs. The training is performed on a single GPU for about 60 hours.

\paragraph{Performance}
The complete simulation-inference pipeline is implemented in CUDA and TensorRT, and the final smoke rendering with neural-predicted lightmaps is integrated into an OpenGL rasterization pass. We use a weakly compressible velocity solver and grid-based smoke advection, employing a $100^3$ velocity grid and a $400^3$ density grid to balance efficiency and fidelity. The computation of velocity and the advection of density take 4.0 ms and 11.5 ms, respectively. Guiding map generation and neural lightmaps prediction are performed at $512^2$ resolution, requiring ~2 ms for guiding-map construction and fixed 1.2 ms for inference. The final rendering stage costs less than 0.3 ms. Our depth-based shadowing adds only 0.8 ms of overhead for rigid-body shadows cast onto the smoke.

\begin{figure*}[ht!]
\centering
\newcommand{\figcap}[1]{\begin{minipage}{0.249\linewidth}\centering#1\end{minipage}}
\includegraphics[trim = 200 120 200 120, clip, width=0.249\linewidth]{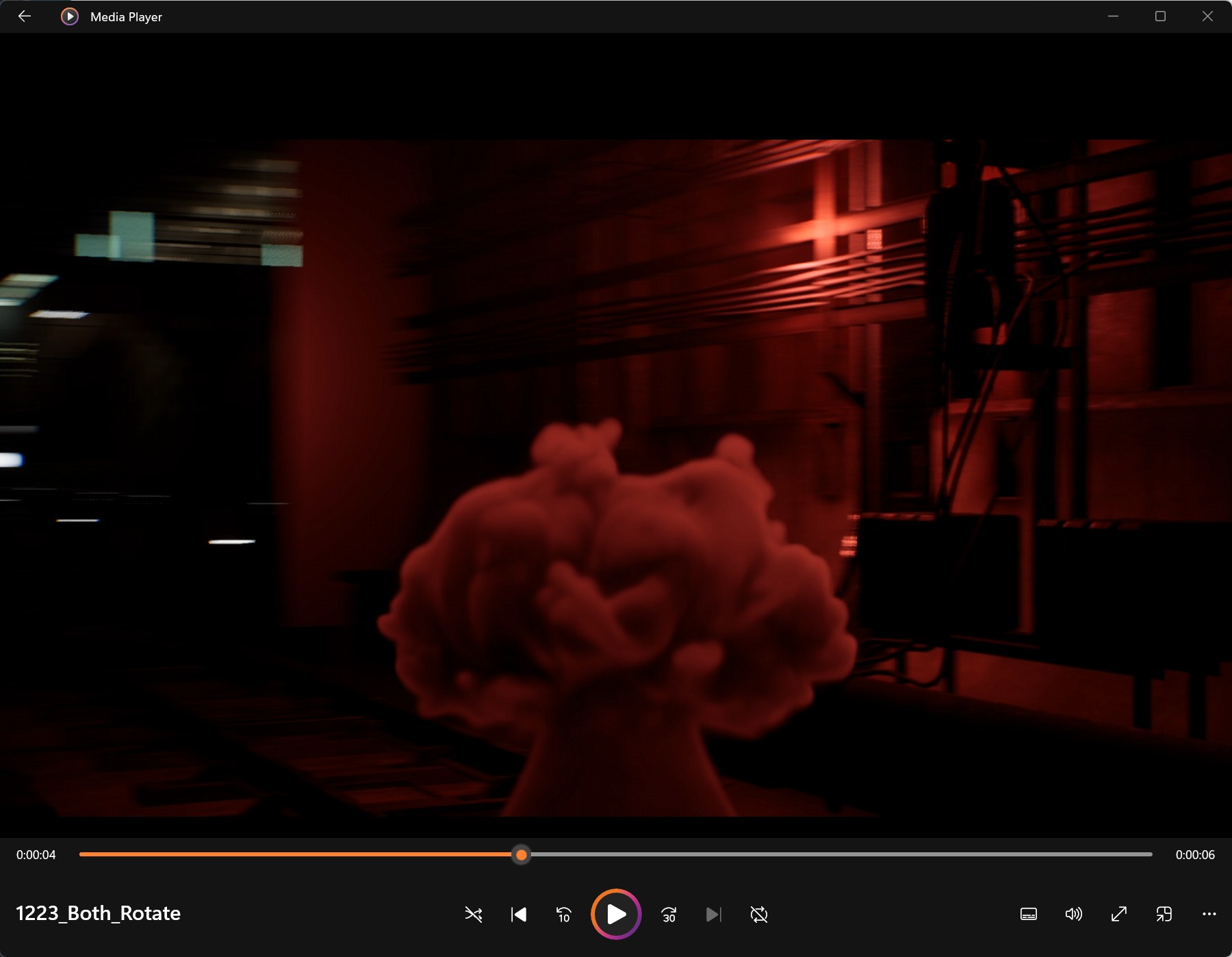}\hfill
\includegraphics[trim = 200 120 200 120, clip, width=0.249\linewidth]{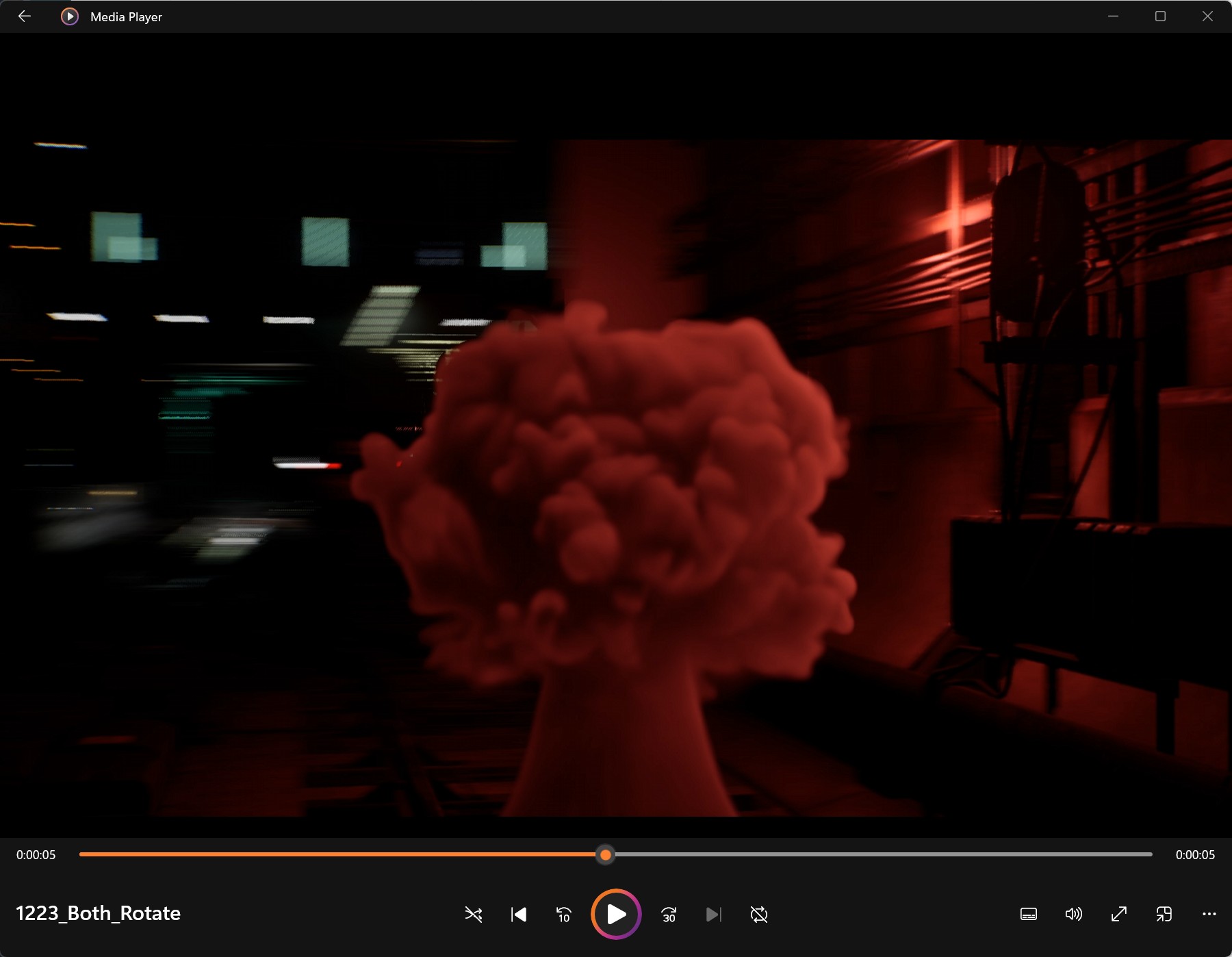}\hfill
\includegraphics[trim = 200 120 200 120, clip, width=0.249\linewidth]{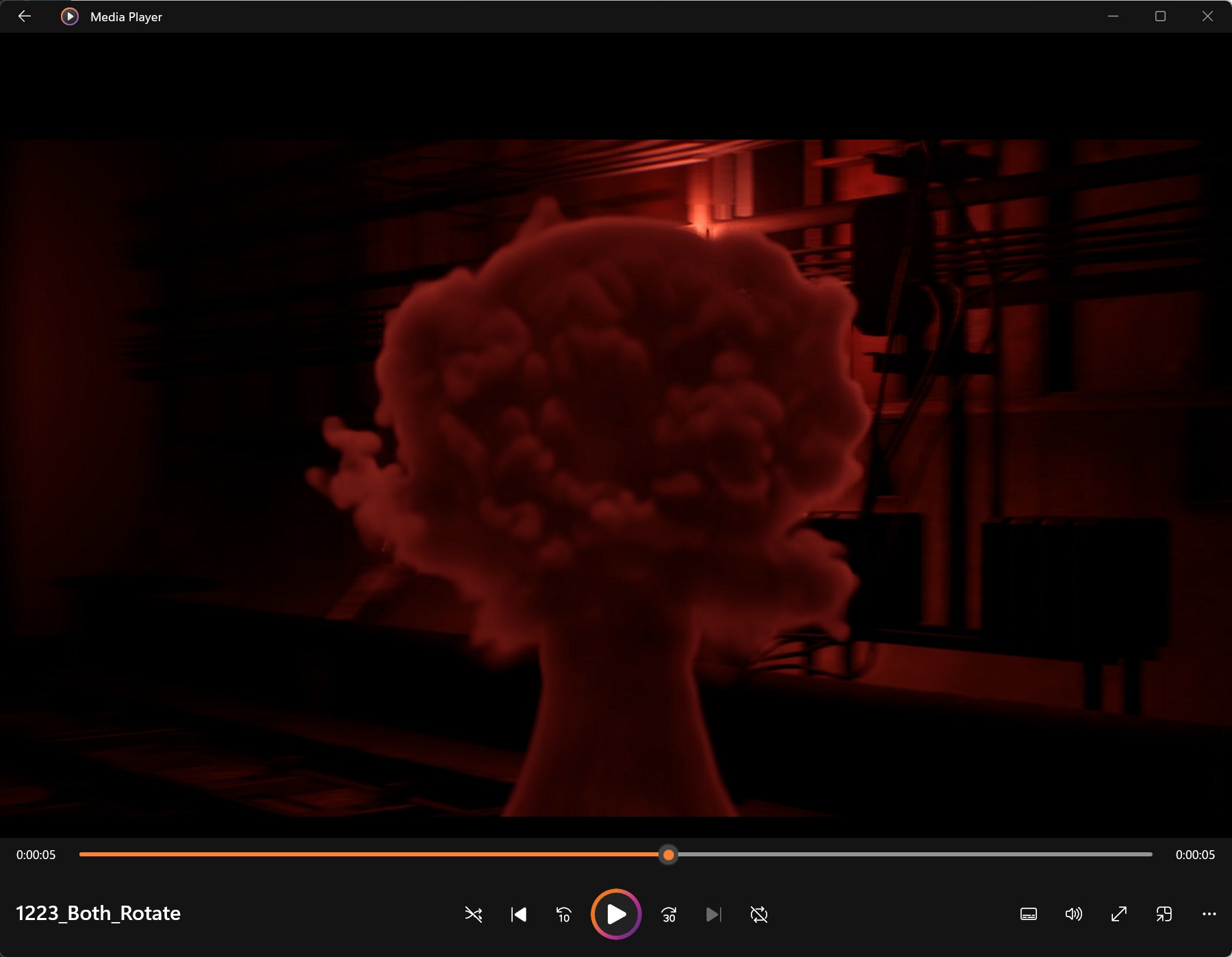}\hfill
\includegraphics[trim = 200 120 200 120, clip, width=0.249\linewidth]{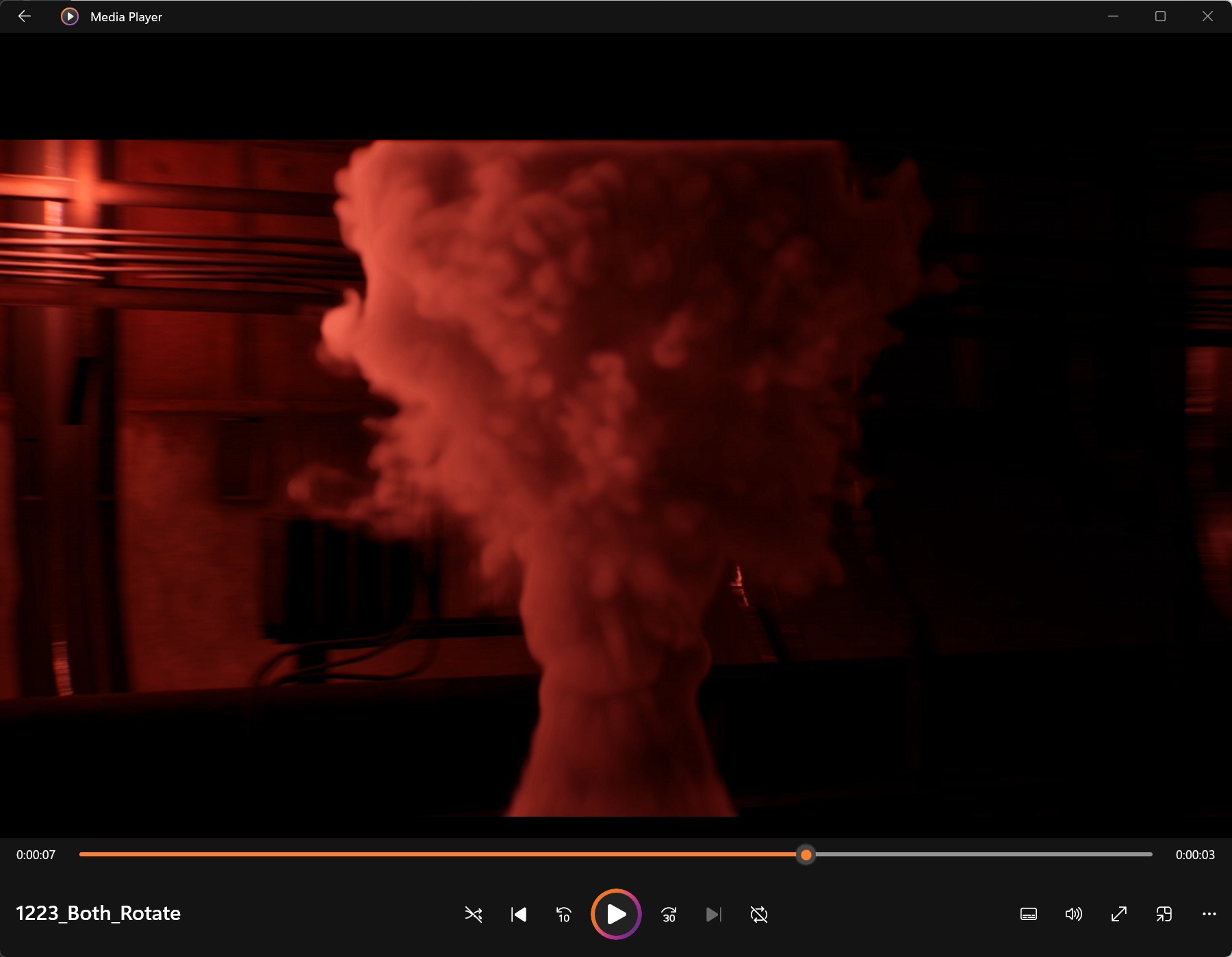}
\caption{\label{fig:plume} Our approach can handle dynamic smoke under a moving camera and be integrated seamlessly into Unreal Engine~\cite{unrealengine}.} \vspace{-1em}
\Description{}
\end{figure*}

\paragraph{Comparison with Volumetric ReSTIR~\cite{Lin2021}}
We compare our method with state-of-the-art fast volumetric rendering techniques, volumetric ReSTIR~\cite{Lin2021}, as shown in \autoref{fig:comp}. With only 1 sample per pixel (SPP), volumetric ReSTIR exhibits significant noise. Also, since it evaluates full light paths via standard tracking, its rendering time exceeds 10 ms even at this low sampling rate. While OptiX temporal denoising can substantially improve visual quality, it adds additional ~2 ms overhead. Beyond performance, our results preserve more structural detail and achieve higher Peak Signal-to-Noise Ratio (PSNR) and lower Mean Squared Error (MSE).

\paragraph{Comparison with MRPNN~\cite{Hu2023}}
Compared to the learned real-time method, our approach eliminates costly precomputation. MRPNN requires constructing a multi-level 3D mipmap hierarchy for each simulation frame, which takes over 170 ms for a $400^3$ density grid in our profiling. When the light direction or position changes, MRPNN must also recompute per-voxel transmittance fields, adding another ~4 ms. Its inference and final rendering take an additional 4 ms and 0.5 ms, respectively. In contrast, our method requires only ~3 ms for inference and 0.5 ms for rendering, thanks to a smaller network and predicting 2D lightmaps rather than full 3D light-scattering fields. More importantly, MRPNN’s shadow prediction is less accurate, resulting in overly bright images and, correspondingly, lower PSNR and higher MSE.

\paragraph{Ablation Study on Guiding Light Directions}
We use three illumination directions, front, top, and bottom, when generating the guiding map for smoke illumination. We validate this choice by comparing it with alternative configurations, including a front-only configuration and a front–left–right combination. As shown in~\autoref{fig:directionstudy}, while including left and right lighting yields reasonably accurate inference results, the front–top–bottom configuration consistently achieves the best overall results.

\paragraph{Ablation Study on Density Variation} 
We evaluate our method under varying density scales, as shown in~\autoref{fig:density100}. Although the network is trained only on inlet densities of 0.1, we test on density fields scaled by $0.5\times$ and $2.0\times$. While performance decreases slightly, dropping a few PSNR points due to the distribution shift, the results remain stable, with overall PSNR values above 30. Even at $2.0\times$ density, our method continues to outperform both denoised ReSTIR and MRPNN, demonstrating stronger robustness to density variation, as illustrated in \autoref{fig:density}.

\paragraph{Ablation Studies on Losses}
We further perform an ablation study of all loss terms and network design choices on an unseen animation sequence simulated by our LBM solver, as shown in \autoref{fig:ablation}, reporting average, minimum, and maximum peak PSNR and MSE. Each component proves essential for achieving high-quality results. In particular, the channel adapters significantly improve accuracy by leveraging partial structural similarities among lightmaps from different orientations. Using sRGB space helps preserve fine details, especially in darker regions, where numerical precision is limited. The flow loss enforces temporal coherence across frames, while the perceptual loss maintains high-level structural features that are difficult to capture with pixel-wise losses alone.


\paragraph{Jet Flow}`
\autoref{fig:jetflow} presents a jet-flow example simulated with an LBM solver. The proposed neural lightmap method faithfully illuminates the dense, fast-moving smoke while preserving fine-scale vortex structures. By rotating both the camera and the light direction, we demonstrate strong spatial and temporal coherence; additional results are provided in the supplemental video.
Our framework also supports smoke–obstacle interactions. As shown in \autoref{fig:bunny}, a rigid bunny is immersed in the jet flow, and our depth-based screen-space shadow approximation produces visually coherent shadows cast from the object onto the smoke with negligible performance cost.
Finally, \autoref{fig:sphere} shows an additional example in which a moving sphere interacts with the jet flow. Leveraging our real-time fluid simulation and lightweight neural lightmaps, the entire simulation-and-rendering pipeline runs at 10 ms per frame, with 7 ms for simulation and 3 ms for rendering, respectively.

\paragraph{Explosion and Generalization to Out-of-distribution Data}
The emission term is crucial for explosion rendering because of the self-luminous nature of the phenomenon. To generate training data for the emissive channel, we use Houdini six-way lightmap toolkit~\cite{Houdini}. Specifically, the Houdini Pyro Bake Volume geometry node converts the simulated smoke density into a specialized scatter volume, which serves as a one-dimensional emissive texture. During rendering, the values of this 1D emissive texture are remapped to RGB colors using Houdini default fire intensity mapping function, producing the final explosion appearance. As shown in~\autoref{fig:explosion}, we evaluate our method on a sequence of pre-simulated unseen volumetric datasets obtained from a third-party source. By applying our network directly to these density fields, we demonstrate that it can accurately predict the emissive channel of six-way lightmaps, enabling real-time approximation of complex volumetric lighting effects that include both scattering and emission. \autoref{fig:third} demonstrates our method applied to a sequence of pre-simulated volumetric explosion data directly downloaded from a third-party website to demonstrate our method is capable of applying to out-of-distribution data.

\paragraph{Demos in Unreal Engine~\cite{unrealengine}}
We first demonstrate our method using a smoke plume example in a subway station scene, shown in \autoref{fig:plume}. By illuminating the plume with a red point light, this example highlights the effectiveness of our approach in handling dynamic smoke under a moving camera. The entire pipeline integrates seamlessly with an industrial game engine, underscoring its practicality for real-time applications. We further demonstrate interactive adjustment of light color and position in~\autoref{fig:interactive}, the whole sequence can be found in the supplemental video. Finally, we present a chimney smoke example in~\autoref{fig:teaser}. Compared with the UE-built-in Niagara system using sprite-based particle effects, our method produces substantially more realistic smoke appearance through lightmap lighting. Moreover, while standard six-way lightmaps enable camera-oriented billboards, they often introduce noticeable artifacts for asymmetric smoke shapes.

\paragraph{Discussion}
Our method relies on a screen-space guiding map to predict lightmaps, enabling high efficiency but also limiting our ability to capture fully accurate shadowing. To address this, we use a simple yet effective technique to approximate the shadows cast by objects onto the smoke. Since accurate shadow requires ray marching from scene geometry toward the light source, we omit this in our framework because such effects are commonly omitted in real-time applications for performance reasons.

%% file: conclusion.tex
\section{Conclusion}

We have introduced a neural lightmaps framework for real-time smoke rendering that learns light scattering from a 2D guiding map, providing a practical alternative for real-time volume rendering methods. By leveraging a UNet with specialized channel adapters, our method predicts six-way lightmaps and transparency maps from screen-space cues, enabling seamless integration into existing game engine pipelines.
Our approach achieves high visual fidelity while maintaining interactive performance, supporting dynamic view directions, object–smoke interactions, and real-time interactivity. 

\paragraph{Limitations}
As shown in~\autoref{fig:vdb_bunny}, our method achieves higher PSNR and MSE than prior work~\cite{Hu2023,Lin2021}, but it still struggles to predict shadowing for data that deviates significantly from the training distribution, revealing a limitation in generalization that remains an avenue for future work. Another limitation is that shadowing cast by external objects is handled with an ad-hoc approximation targeting visual plausibility rather than physical accuracy for real-time applications. A more principled approach, such as computing object-cast shadow volumes throughout the medium rather than relying solely on surface projection, offers a promising direction for improving fidelity.